\documentclass[nofootinbib,aps,preprint,preprintnumbers,superscriptaddress]{revtex4}
\usepackage[dvipdfmx]{graphicx}
\usepackage{amssymb}
\usepackage{amsmath}              
\usepackage{mathtools}
\usepackage{cases}              
\usepackage{bm}
\usepackage{color}

\newcommand{\VEV}[1]{\langle #1 \rangle}

\newcommand{\tr}{\mbox{tr}}
\newcommand{\fh}{\mathfrak{h}}
\newcommand{\fg}{\mathfrak{g}}

\newcommand{\diagentry}[1]{\mathmakebox[1.8em]{#1}}
\newcommand{\xddots}{%
  \raise 4pt \hbox {.}
  \mkern 6mu
  \raise 1pt \hbox {.}
  \mkern 6mu
  \raise -2pt \hbox {.}
}

\def\l{\left}
\def\r{\right}

\begin{document}

\preprint{KUNS-2755}
\title{
Symmetry and geometry in \\
generalized Higgs effective field theory \\
-- Finiteness of oblique corrections v.s. perturbative unitarity --
}
\author{Ryo Nagai}
\email[E-mail: ]{rnagai@icrr.u-tokyo.ac.jp}
\affiliation{
Institute for Cosmic Ray Research (ICRR), The University of Tokyo, Kashiwa, Chiba 277-8582, Japan
}
\affiliation{
Department of Physics, Tohoku University, Sendai, Miyagi
980-8578, Japan
}
\author{Masaharu Tanabashi}
\email[E-mail: ]{tanabash@eken.phys.nagoya-u.ac.jp}
\affiliation{
Department of Physics,
Nagoya University, Nagoya 464-8602, Japan
}
\affiliation{
Kobayashi-Maskawa Institute for the Origin of Particles and the
Universe, 
\\ Nagoya University, Nagoya 464-8602, Japan
}
\author{Koji Tsumura}
\email[E-mail: ]{ko2@gauge.scphys.kyoto-u.ac.jp}
\affiliation{
Department of Physics,
Kyoto University, Kyoto 606-8502, Japan
}
\author{Yoshiki Uchida}
\email[E-mail: ]{uchida@eken.phys.nagoya-u.ac.jp}
\affiliation{
Department of Physics,
Nagoya University, Nagoya 464-8602, Japan
}

\date{\today}

\begin{abstract} 
We formulate a generalization of Higgs effective field theory (HEFT) including arbitrary number of extra neutral and charged Higgs bosons
(generalized HEFT, GHEFT) to describe non-minimal electroweak symmetry
breaking models. 
Using the geometrical form of the GHEFT Lagrangian, which can be
regarded as a nonlinear sigma model on a scalar manifold, 
it is shown that 
the scalar boson scattering amplitudes are described in terms of the
Riemann curvature tensor (geometry) of the scalar manifold and the covariant derivatives
of the potential.
The coefficients of the one-loop divergent terms 
in the oblique correction parameters $S$ and
$U$ can also be written in terms of the Killing vectors (symmetry)
and the Riemann curvature tensor (geometry). 
It is found that perturbative unitarity of the scattering amplitudes 
involving the Higgs bosons and the longitudinal gauge bosons 
demands the flatness of the scalar manifold. 
The relationship between the finiteness of the electroweak oblique 
corrections and perturbative unitarity of the scattering amplitudes is also clarified 
in this language: we verify that once the tree-level unitarity is ensured, 
then the one-loop finiteness of the oblique correction parameters $S$ and $U$
is automatically guaranteed. 
\end{abstract}

\maketitle

\section{Introduction}

What is the origin of the electroweak symmetry breaking (EWSB)? 
In the standard model (SM) of particle physics, 
the EWSB is caused by a vacuum expectation value of a complex 
scalar field (SM Higgs field), 
which linearly transforms under the $SU(2)_W\times U(1)_Y$ 
electroweak gauge symmetry. 
The Higgs sector of the SM is constructed to be minimal,
as it includes only a scalar boson (SM Higgs boson) and 
three would-be Nambu-Goldstone bosons eaten by massive gauge bosons
after the EWSB\@.
There are no cousin particles of Higgs in the SM.
The scalar particle discovered by the ATLAS and CMS experiments in 2012 
with the mass of 125\,GeV \cite{Aad:2012tfa,Chatrchyan:2012ufa}
can now be successfully interpreted as the SM(-like) Higgs boson.

The Higgs sector in the SM, however, does not ensure the stability of the EWSB scale against quantum corrections. 
In other words, the SM itself cannot explain why the EWSB scale is an order of 100\,GeV, 
much smaller than its cutoff scale such as Planck 
(or Grand Unification) scale. 
The SM Higgs sector is therefore inherently incomplete.
It should be extended. 
Many extensions/generalizations of the SM Higgs sector, 
such as Two Higgs Doublet Model \cite{Haber:1978jt,Deshpande:1977rw,Georgi:1978xz,Donoghue:1978cj,Abbott:1979dt,McWilliams:1980kj,Gunion:1984yn,Branco:2011iw,Cheon:2012rh,Craig:2012vn,Chang:2012ve,Bai:2012ex,Ferreira:2012nv,Chang:2012zf,Chen:2013kt,Celis:2013rcs,Grinstein:2013npa,Chen:2013rba,Craig:2013hca,Kanemura:2013eja,Ferreira:2014naa,Kanemura:2014bqa}, 
Composite Higgs Models \cite{Kaplan:1983fs,Kaplan:1983sm,Georgi:1984ef,Georgi:1984af,Dugan:1984hq,Contino:2003ve,Agashe:2004rs,Mrazek:2011iu,DeCurtis:2018iqd,DeCurtis:2018zvh}, Georgi-Machacek Model \cite{Georgi:1985nv,Chanowitz:1985ug,Gunion:1989ci,Gunion:1990dt}, 
etc.,
have been proposed.
The 125GeV Higgs boson accompanies extra Higgs particles in these scenarios.

The Effective Field Theory (EFT) approach is widely used to study
these beyond-SM (BSM) physics in a model independent manner.
The physics below 1TeV can be described by the Standard Model 
Effective Field Theory (SMEFT) \cite{Buchmuller:1985jz,Grzadkowski:2010es,De Rujula:1991se,Hagiwara:1992eh,Hagiwara:1993ck,Hagiwara:1993qt,Alam:1997nk,Elias-Miro:2013mua,Barger:2003rs,Kanemura:2008ub,Corbett:2012dm,Corbett:2012ja,Grojean:2013kd,Elias-Miro:2013gya,Corbett:2013pja,Mebane:2013cra,Belanger:2013xza,Lopez-Val:2013yba,Jenkins:2013zja,Jenkins:2013wua,Alonso:2013hga,Boos:2013mqa,Ellis:2014dva,Ellis:2014jta,Falkowski:2014tna,Henning:2014wua,Contino:2016jqw,Ellis:2018gqa}, which parametrizes the BSM contributions 
using the coefficients of SM field higher dimensional operators.
The SMEFT is successful if the BSM particles are much heavier than 1TeV
and they decouple from the low energy physics.
The SMEFT cannot be applied, however, if the heavy BSM particles do not decouple from the low energy physics.
The Higgs Effective Field Theory (HEFT) \cite{Feruglio:1992wf,Burgess:1999ha,Giudice:2007fh,Grinstein:2007iv,Alonso:2012px,Buchalla:2012qq,Azatov:2012bz,Contino:2013kra,Jenkins:2013fya,Buchalla:2013rka,Buchalla:2013eza,Alonso:2014rga,Guo:2015isa,Buchalla:2015qju,Alonso:2017tdy,Buchalla:2017jlu,Buchalla:2018yce} should be 
applied instead.
These existing EFTs cannot be applied if there exist BSM 
particles lighter than 1TeV.
We should include these BSM particles explicitly in the EFT approach.

In this paper, we propose a generalization of HEFT (GHEFT) for this
purpose.
As in the HEFT, GHEFT is based on the electroweak chiral perturbation
theory (EWChPT) \cite{Appelquist:1980vg,Appelquist:1980ae,Longhitano:1980iz,Longhitano:1980tm,Appelquist:1993ka,Appelquist:1994qz}.
In GHEFT, the BSM particles, as well as the 125GeV Higgs boson, are 
introduced as matter particles in the Callan-Coleman-Wess-Zumino (CCWZ)
construction \cite{Coleman:1969sm,Callan:1969sn,Bando:1987br} of EWChPT\@.

Note that the longitudinal gauge boson scattering amplitudes exceed
perturbative unitarity limits at high energy in the EWChPT.
The GHEFT couplings should satisfy special conditions, known as 
the unitarity sum rules \cite{Gunion:1990kf,Csaki:2003dt,SekharChivukula:2008mj}, to keep 
the amplitudes perturbative in the high energy scatterings,
if the model is considered to be ultraviolet (UV) complete.
We also note that the EWChPT is not renormalizable.
The UV completed GHEFT couplings should satisfy the finiteness conditions 
in order to cancel these UV divergences.

The GHEFT can also be described in a geometrical language using
the scalar manifold metric, as discussed in Refs. \cite{Alonso:2015fsp,Alonso:2016oah} in the HEFT 
context.
We point out that both the scalar scattering amplitudes and the
one-loop UV divergences in the electroweak oblique correction parameters
$S$ and $U$ \cite{Peskin:1990zt} are described by using the Riemann curvature tensor
(geometry) and the Killing vectors (symmetry) of the scalar manifold.
Therefore, both the unitarity sum rules and the oblique correction 
finiteness conditions are described in terms of the geometry and the symmetry.
We find that the perturbative unitarity is ensured by the flatness 
of the scalar manifold (vanishing Riemann curvature).
We also find that the divergences in the oblique correction parameters
($S$ and $U$ parameters) are canceled if a subset of the 
perturbative unitarity conditions and the $SU(2)_W\times U(1)_Y$ gauge symmetry 
are satisfied.
These findings generalize our previous observation \cite{Nagai:2014cua} which relates
 the perturbative unitarity to the one-loop finiteness of the oblique correction parameters{\footnote{
Possible relations between the unitarity and the renormalizabilty have also been investigated in gravity models. See Refs.~\cite{Fujimori:2015wda,Fujimori:2015mea,Fujimori:2016rrc,Abe:2017abx,Abe:2018rwb}. 
 }}.

This paper is organized as follows: 
in \S.~\ref{sec:model} we introduce the GHEFT Lagrangian at its lowest
order (${\cal O}(p^2)$).
We investigate the scalar boson scattering amplitudes 
in \S.~\ref{sec:Amplitude}.
\S.~\ref{sec:one-loop-gaugeless} and \S.~\ref{sec:oblique}
are for one-loop computations with and without the gauge boson 
contributions.
The relationship between the perturbative unitarity and the
one-loop finiteness of the oblique correction parameters
is clarified in \S.~\ref{sec:vs}.
We conclude in \S.~\ref{sec:sum}.

\section{Generalized HEFT Lagrangian of 
$SU(2)_W \times U(1)_Y \to U(1)_{\rm em}$}
\label{sec:model}

The electroweak chiral perturbation theory (EWChPT) \cite{Appelquist:1980vg,Appelquist:1980ae,Longhitano:1980iz,Longhitano:1980tm,Appelquist:1993ka,Appelquist:1994qz} provides a 
systematic framework to describe the low energy phenomenologies
of the electroweak symmetry breaking physics.
It utilizes the electroweak chiral Lagrangian method for 
parametrizing the non-decoupling corrections,  
which appear ubiquitously in models with strongly interacting 
electroweak symmetry breaking sector.
Although the original version of the EWChPT was constructed to 
be a Higgsless theory \cite{Appelquist:1980vg,Longhitano:1980iz,Longhitano:1980tm,Appelquist:1980ae,Csaki:2003dt,SekharChivukula:2008mj,Cacciapaglia:2004rb,Foadi:2004ps,Casalbuoni:2005rs,Cacciapaglia:2005pa,Foadi:2005hz,Chivukula:2005bn,Chivukula:2005xm,Chivukula:2006cg,Abe:2008hb,Abe:2011sv}, 
after the discovery of the 125GeV Higgs particle, the EWChPT is 
extended to the Higgs Effective Field Theory (HEFT) \cite{Feruglio:1992wf,Burgess:1999ha,Giudice:2007fh,Grinstein:2007iv,Alonso:2012px,Buchalla:2012qq,Azatov:2012bz,Contino:2013kra,Jenkins:2013fya,Buchalla:2013rka,Buchalla:2013eza,Alonso:2014rga,Guo:2015isa,Buchalla:2015qju,Alonso:2017tdy,Buchalla:2017jlu,Buchalla:2018yce}, incorporating
the 125 GeV Higgs particle $h$ as a neutral spin-0 matter particle 
in the electroweak chiral Lagrangian.
Introducing functions $\mathcal{F}(h)$ and $V(h)$, which parametrize
the phenomenological properties of the 125GeV Higgs, 
the HEFT provides a systematic description for a neutral 
spin-0 particle in the electroweak symmetry breaking sector, 
including the one-loop radiative corrections 
\cite{Buchalla:2012qq,Alonso:2012px,Contino:2013kra,Jenkins:2013fya,Buchalla:2013rka,Buchalla:2013eza,Alonso:2014rga,Guo:2015isa,Buchalla:2015qju,Alonso:2017tdy,Buchalla:2017jlu,Buchalla:2018yce}.  
It can parametrize the low energy properties of the 125GeV Higgs 
particle in the strongly interacting model context, as well as 
weakly interacting model context.

We need to generalize the HEFT further (generalized HEFT, GHEFT), if we want to introduce
extra Higgs particles other than the discovered 125GeV Higgs 
particle.  
It is not trivial to introduce non-singlet extra particles
in the EWChPT, however, 
since the electroweak gauge symmetry $SU(2)_W \times U(1)_Y$ is 
realized nonlinearly in the EWChPT\@.
The interaction Lagrangian
needs to be arranged carefully to make the theory invariant 
under the electroweak gauge symmetry $SU(2)_W\times U(1)_Y$.

These extra non-singlet Higgs particles can be regarded as 
matter particles in the EWChPT Lagrangian context.   
The Callan-Coleman-Wess-Zumino (CCWZ) formulation \cite{Coleman:1969sm,Callan:1969sn,Bando:1987br} provides
an ideal framework for the concrete construction of the 
matter particle interaction Lagrangian in a manner consistent 
with the nonlinear sigma model symmetry structure. 
See, {\it{e.g.}},  Refs.~\cite{Ecker:1988te, Alboteanu:2008my} for earlier studies on these non-singlet matter particles in QCD chiral perturbation theory and EWChPT, respectively.

In this section, we apply the CCWZ formulation 
for the construction of the GHEFT Lagrangian.

\subsection{
Electroweak chiral Lagrangian
}

For simplicity, in this subsection, 
we consider the EWChPT Lagrangian 
in the gaugeless limit, {\it{i.e.}}, $g_W=g_Y=0$.
The couplings with the electroweak gauge fields will be 
introduced in \S.~\ref{sec-ewgauge-fields}.
The electroweak symmetry $G=[SU(2)_W \times U(1)_Y]$ is broken 
spontaneously to the $H=U(1)_{\rm em}$ symmetry in the SM
Higgs sector.
The most general scalar sector Lagrangian consistent with
the symmetry breaking structure 
$G/H=[SU(2)_W\times U(1)_Y]/U(1)_{\rm em}$ can be constructed
as the CCWZ nonlinear sigma model Lagrangian on the coset space $G/H$. 
The coset manifold $G/H=[SU(2)_W \times U(1)_Y]/U(1)_{\rm em}$
is coordinated by the Nambu-Goldstone (NG) boson fields $\pi^a$ ($a=1,2,3$) as
\begin{equation}
  \xi_W(x) = \exp\biggl( i \sum_{a=1,2} \pi^a(x) \dfrac{\tau^a}{2} \biggr) 
\, , 
\label{eq:coset-coordinate1}
\end{equation}
\begin{equation}
  \xi_Y(x) = \exp\biggl( i \pi^3(x) \dfrac{\tau^3}{2} \biggr)  
\, ,
\label{eq:coset-coordinate2}
\end{equation}
with $\tau^a~(a=1,2,3)$ being Pauli spin matrices.
Under the $G=[SU(2)_W \times U(1)_Y]$ transformation,
\begin{equation}
  \fg_W \in SU(2)_W \, , 
  \qquad 
  \fg_Y \in U(1)_Y \, ,
\end{equation}
these NG boson fields transform as
\begin{equation}
  \xi_W(x) \to 
  \xi'_W(x) = \fg_W \, \xi_W(x) \, \fh^\dag(\pi, \fg_W, \fg_Y) \, , 
\label{eq:transf1}
\end{equation}
\begin{equation}
  \xi_Y(x) \to
  \xi'_Y(x) = \fh(\pi,\fg_W, \fg_Y) \,  \xi_Y(x) \, \fg_Y^\dag \, .
\label{eq:transf2}
\end{equation}
Here $\fh(\pi, \fg_W, \fg_Y)$ is an element of the unbroken group $H$,
which is determined to pull-back the coset space coordinates to their
original forms (\ref{eq:coset-coordinate1}) and 
(\ref{eq:coset-coordinate2}).
Note that the $H$ transformation $\fh(\pi, \fg_W, \fg_Y)$ depends
not only on the $SU(2)_W$ and $U(1)_Y$ elements 
$\fg_{W}$ and $\fg_{Y}$, but also on the NG boson fields $\pi(x)$.
The NG boson fields $\pi^a$ ($a=1,2,3$) therefore transform nonlinearly 
under the $G$ symmetry.

It is useful to introduce objects called Maurer-Cartan (MC) one-forms
$\alpha_{\perp \mu}^a$ ($a=1,2,3$) defined as
\begin{align}
  \alpha_{\perp \mu}^a 
  &=  \tr\l[ \dfrac{1}{i} \xi_W^\dag (\partial_\mu \xi_W) \tau^a \r]
\, , 
\qquad (a=1,2)\\
\intertext{and}
  \alpha_{\perp \mu}^3 
  &=  
  \tr\l[ \dfrac{1}{i} \xi_W^\dag (\partial_\mu \xi_W) \tau^3 \r]
 +\tr\l[ \dfrac{1}{i} (\partial_\mu \xi_Y) \xi_Y^\dag \tau^3 \r] \, .
\label{eq:mc-oneform3}
\end{align}
Although the NG boson fields $\pi$ transform nonlinearly,
these MC one-forms transform homogeneously, {\em i.e.,}
\begin{align}
  \sum_{a=1,2}\alpha_{\perp\mu}^a \dfrac{\tau^a}{2}
  &\to 
  \fh(\pi, \fg_W, \fg_Y) \, 
  \left( \sum_{a=1,2}\alpha_{\perp\mu}^a \dfrac{\tau^a}{2} \right) \, 
  \fh^\dag(\pi, \fg_W, \fg_Y) \, ,
\label{eq:mc-form-trans1}
\\
  \alpha_{\perp\mu}^3 \dfrac{\tau^3}{2}
  &\to 
  \fh(\pi, \fg_W, \fg_Y) \, 
  \left( \alpha_{\perp\mu}^3 \dfrac{\tau^3}{2} \right) \, 
  \fh^\dag(\pi, \fg_W, \fg_Y) \, ,
\label{eq:mc-form-trans2}
\end{align}
under the $G$ symmetry.
We see that the MC one-forms transform as
\begin{equation}
  \alpha_{\perp\mu}^a \to 
  \l[\rho_\alpha(\fh)\r]^a{}_b \alpha_{\perp\mu}^b \, ,
\label{eq:transMC}
\end{equation}
with $\rho_\alpha(\fh)$ being a $3\times 3$ matrix
\begin{equation}
  \rho_\alpha(\fh)
  = \exp\biggl( i \theta_h(\pi, \fg_W, \fg_Y) \, 
    Q_\alpha \biggr) \, , \qquad
  \fh
  = \exp\biggl( i \theta_h(\pi, \fg_W, \fg_Y) \, \dfrac{\tau^3}{2} 
    \biggr)\, .
\end{equation}
In the expression (\ref{eq:transMC}) and hereafter,
summation ${\displaystyle \sum_{b=1,2,3}}$ 
is implied whenever an index $b$ is repeated in a product.
Here the NG boson charge matrix $Q_\alpha$ is defined by
\begin{equation}
  Q_\alpha = 
  \begin{pmatrix}
    \diagentry{-\sigma_2} \\
    & \diagentry{0}
  \end{pmatrix} \, ,
\end{equation}
with $\sigma_2$ being the Pauli spin matrix 
\begin{equation}
  \sigma_2 = 
  \begin{pmatrix}
    0 & -i \\
    +i & 0 
  \end{pmatrix}
  \, .
\end{equation}

It is now straightforward to construct the lowest order 
(${\cal O}(p^2)$) $G$ invariant
Lagrangian of the NG bosons:
\begin{equation}
  {\cal L}_{\pi}
  = \dfrac{1}{2} G_{ab}^{(0)} \alpha_{\perp\mu}^a \, \alpha_{\perp}^{b\mu}
\label{eq:ewchlag1}
\end{equation}
with
\begin{equation}
  G_{ab}^{(0)}
  = \dfrac{1}{4} 
  \begin{pmatrix}
    \diagentry{v^2} \\
    & \diagentry{v^2} \\
    & & \diagentry{v_Z^2}
  \end{pmatrix}
  \, .
\end{equation}
The Lagrangian can be rewritten as
\begin{equation}
  {\cal L}_{\pi}
  = \dfrac{v^2}{4} \tr \l[
       (\partial_\mu U^\dagger ) \, 
       (\partial^\mu U ) 
    \r]  
   -\dfrac{v_Z^2-v^2}{8}
     \tr \l[ U^\dag (\partial_\mu U) \tau^3 \r]  \, 
     \tr \l[ U^\dag (\partial^\mu U) \tau^3 \r]  \, ,
\label{eq:lag2}
\end{equation}
with
\begin{equation}
   U := \xi_W \xi_Y \, .
\end{equation}
It should be emphasized here that $v$ and $v_Z$ (decay constants
of $\pi^{1,2}$ and $\pi^3$) are independently adjustable parameters
in the EWChPT on the $G/H=[SU(2)_W\times U(1)_Y]/U(1)_{\rm em}$
coset space.
Phenomenologically preferred relation
\begin{equation}
  \rho := \dfrac{v^2}{v_Z^2} \simeq 1
\end{equation}
is realized only by a parameter tuning $v \simeq v_Z$ in 
this setup\footnote{
It is possible to introduce custodial symmetry to justify 
the tuning.
The standard model, in fact, possesses the custodial symmetry 
in its gaugeless and Yukawa-less limit.
See, {\it e.g.}, Ref.~\cite{deFlorian:2016spz} for the HEFT
power counting rules and how custodial symmetry violating
terms are organized therein.
We do not introduce the custodial symmetry here, however, since
it is not relevant with the main findings in the present paper.
The restrictions on the GHEFT Lagrangian parameters coming from 
the custodial requirements and its power counting rules will be 
studied in a separate publication.
}.

\subsection{
Matter particles coupled with the electroweak chiral Lagrangian
}

Thanks to the homogeneous transformation properties of the 
MC one-forms (\ref{eq:transMC}), 
matter particles can be introduced easily in the 
CCWZ formulation of the EWChPT Lagrangian (\ref{eq:lag2}).

We consider a set of real scalar matter fields $\phi^I$, 
which transforms homogeneously as
\begin{equation}
  \phi^I \to [\rho_\phi(\fh)]^I{}_J \, \phi^J \, ,
\label{eq:transf3}
\end{equation}
under the unbroken group $H$.
Here $\rho_\phi(\fh)$ stands for a representation matrix 
\begin{equation}
  \rho_\phi(\fh)
  = \exp \biggl( i \theta_h \, Q_\phi \biggr) \, ,
\qquad
  \fh = \exp\l( i \theta_h \dfrac{\tau^3}{2}
        \r) \, ,
\end{equation}
with $Q_\phi$ being a hermitian matrix.
Note here that the $\fh$ transformation depends on the 
NG boson fields $\pi(x)$.
It therefore is a local transformation depending on the 
spacetime point $x$.
If the set of scalar matter particles consists
of $n_N$ species of neutral particles and $n_C$ species of charged
particles, the matrix $Q_\phi$ can be expressed as a 
$(2n_C+n_N)\times (2n_C+n_N)$ matrix
\begin{equation}
  Q_\phi = 
  \begin{pmatrix}
    \diagentry{-q_1\sigma_2} \\
    &\diagentry{\ddots}\\
    &&\diagentry{-q_{n_C} \sigma_2} \\
    &&& \diagentry{0} \\
    &&&& \diagentry{\ddots} \\
    &&&&& \diagentry{0}
  \end{pmatrix} \, .
\end{equation}
Here $q_i$ ($i=1,2,\cdots n_C$) are
the charges of the scalar matter particles.
Since $\fh$ is a local transformation, 
$\partial_\mu \phi^I$ transforms non-homogeneously under $\fh$.
In order to write a kinetic term for the matter field
$\phi^I$, we therefore introduce a covariant derivative 
of the matter 
field $\phi^I$ : 
\begin{equation}
  (\mathcal{D}_\mu \phi)^I 
  = \partial_\mu \phi^I + i {\cal V}^3_\mu \, [Q_\phi]^I{}_J \, \phi^J \, ,
\qquad
(I,J=1,2,\cdots, 2n_C+n_N) \, .
\label{eq:covariant-der-matter1}
\end{equation}
We take the connection ${\cal V}^3_\mu$ as
\begin{align}
  {\cal V}^3_\mu 
  = -  \tr \left[
      \dfrac{1}{i} (\partial_\mu \xi_Y) \xi_Y^\dagger \tau^3 
    \right] 
  + c \alpha_{\perp \mu}^3
\, ,
\label{eq:connection-ccwz1}
\end{align}
with $c$ being an arbitrary constant.
Hereafter we take $c=0$ for simplicity.
The covariant derivative (\ref{eq:covariant-der-matter1})
transforms homogeneously 
\begin{equation}
  (\mathcal{D}_\mu \phi)^I  \to
  [\rho_{\phi}(\fh)]^I{}_J \, (\mathcal{D}_\mu \phi)^J \, ,
\end{equation}
as we designed so in Eq.~(\ref{eq:covariant-der-matter1}).
It is now straightforward to write down an ${\cal O}(p^2)$ 
EWChPT Lagrangian including additional scalar bosons with arbitrary charges:
\begin{equation}
  {\cal L}
  = \dfrac{1}{2} G_{ab} \alpha^a_{\perp \mu} \alpha^{b\mu}_{\perp}
   +G_{aI} \alpha^a_{\perp \mu} ({\cal D}^\mu \phi)^I
   +\dfrac{1}{2} G_{IJ} ({\cal D}_\mu \phi)^I ({\cal D}^\mu \phi)^J 
   - V\, .
\label{eq:ewchlag-with-matter1}
\end{equation}
Here $G_{ab}$, $G_{aI}$, $G_{IJ}$ and $V$ are functions of the scalar
fields $\phi^I$.
Also, 
$G_{ab}$, $G_{aI}$ and $G_{IJ}$ transform homogeneously as multiplets
of corresponding representations.
They satisfy\footnote{
Eqs.~(\ref{eq:metric-conditions1})  
and (\ref{eq:potential-conditions1})
are understood to be the tree-level matching conditions between
GHEFT and EWChPT.
They may be modified beyond the tree-level. 
See, {\it e.g.,} Refs.\cite{Tanabashi:1993np,Rosell:2005ai}.
}
\begin{equation}
  G_{ab} \biggr|_{\phi=0} = G_{ab}^{(0)} \, ,
  \quad
  G_{aI}  \biggr|_{\phi=0} = 0 \, , 
  \quad
  G_{IJ}  \biggr|_{\phi=0} = \delta_{IJ} \, ,
\label{eq:metric-conditions1}
\end{equation}
and
\begin{equation}
  \dfrac{\partial}{\partial \phi^I} V \biggr|_{\phi=0} = 0 \, , 
  \quad
  \dfrac{\partial}{\partial \phi^J} 
  \dfrac{\partial}{\partial \phi^I} 
  V \biggr|_{\phi=0} = M_I^2 \delta_{IJ} \, ,
\label{eq:potential-conditions1}
\end{equation}
with $M_I$ being the $\phi^I$ boson mass.
The second and the third conditions in Eq.~(\ref{eq:metric-conditions1}) 
can be achieved by redefining the scalar field $\phi^I$ in the 
Lagrangian.
The first condition in Eq.~(\ref{eq:metric-conditions1}) ensures 
that the extended Lagrangian (\ref{eq:ewchlag-with-matter1}) 
reproduces the lowest order EWChPT Lagrangian (\ref{eq:ewchlag1}) in 
the absence of Higgs particles $\phi^I$. The stability around the vacuum $\phi=0$ is guaranteed by the 
conditions (\ref{eq:potential-conditions1}).

\subsection{Electroweak gauge fields}
\label{sec-ewgauge-fields}
It is easy to introduce the electroweak gauge fields $W^a_\mu$ ($a=1,2,3$)
and $B_\mu$ in our EWChPT Lagrangian (\ref{eq:ewchlag-with-matter1}).
When the gauge coupling is switched on, 
we just need to replace the derivatives $\partial_\mu\xi_W^{}$ and 
$\partial_\mu\xi_Y^{}$ by the covariant derivatives:
\begin{align}
D_\mu\xi_W^{}&=\partial_\mu \xi_W^{}-ig_W^{} 
W^a_\mu\frac{\tau^a}{2}\xi_W^{}
\label{eq:cov-der-xi}
\, 
,\\
D_\mu\xi_Y^{}&=\partial_\mu \xi_Y^{}+ig_Y^{} \xi_Y^{} B_\mu\frac{\tau^3}{2}
\, ,
\end{align}
with $g_W^{}$ and $g_Y^{}$ being the 
$SU(2)_W$ and $U(1)_Y$ gauge coupling strengths, respectively. 

The lowest order ($\mathcal{O}(p^2)$) GHEFT Lagrangian is therefore 
\begin{align}
  {\cal L}
  &= \dfrac{1}{2} G_{ab} \hat{\alpha}^a_{\perp \mu} \hat{\alpha}^{b\mu}_{\perp}
   +G_{aI} \hat{\alpha}^a_{\perp \mu} ({\cal D}^\mu \phi)^I
   +\dfrac{1}{2} G_{IJ} ({\cal D}_\mu \phi)^I ({\cal D}^\mu \phi)^J 
   - V
  \nonumber\\
  & \qquad
   - \dfrac{1}{4} W^a_{\mu\nu} W^{a\mu\nu} -\dfrac{1}{4} B_{\mu\nu} B^{\mu\nu} 
\, ,
\label{eq:ewchlag-with-matter2}
\end{align}
with
\begin{align}
  \hat{\alpha}^a_{\perp \mu}
  &= \tr\left[ \dfrac{1}{i} \xi_W^\dagger (\partial_\mu \xi_W) \tau^a \right]
    -g_W \tr\left[ \xi_W^\dagger W^b_\mu \dfrac{\tau^b}{2} \xi_W  \tau^a \right] 
\, , \qquad   (a=1,2)
  \\
\intertext{and}
  \hat{\alpha}^3_{\perp \mu}
  &= \tr\left[ \dfrac{1}{i} \xi_W^\dagger (\partial_\mu \xi_W) \tau^3 \right]
     +\tr\left[ \dfrac{1}{i} (\partial_\mu \xi_Y) \xi_Y^\dagger   \tau^3 \right]
    -g_W \tr\left[ \xi_W^\dagger W^b_\mu \dfrac{\tau^b}{2} \xi_W  \tau^3 \right]
    +g_Y B_\mu\, .
     & 
\end{align}
We define the covariant derivative of the matter fields
$({\cal D}_\mu \phi)^I$
\begin{equation}
  ({\cal D}_\mu \phi)^I
  = \partial_\mu \phi^I + i \hat{\cal V}^3_\mu [Q_\phi]^I{}_J \phi^J \, ,
\end{equation}
with
\begin{equation}
  \hat{\cal V}^3_\mu = - \tr\left[ 
    \dfrac{1}{i} (\partial_\mu \xi_Y)\xi_Y^\dagger \tau^3 \right]
  -g_Y B_\mu\,. 
\end{equation}

It should be noted that the GHEFT 
Lagrangian (\ref{eq:ewchlag-with-matter2}) reproduces 
HEFT Lagrangian \cite{Feruglio:1992wf,Burgess:1999ha,Giudice:2007fh,Grinstein:2007iv,Alonso:2012px,Buchalla:2012qq,Azatov:2012bz,Contino:2013kra,Jenkins:2013fya,Buchalla:2013rka,Buchalla:2013eza,Alonso:2014rga,Guo:2015isa,Buchalla:2015qju,Alonso:2017tdy,Buchalla:2017jlu,Buchalla:2018yce} for $n_N=1$ and $n_C=0$.
Here $\phi^{I=h}$ stands for the 125\, GeV Higgs boson field.
In the HEFT, $G_{aI}$ and $G_{IJ}$ are taken as
\begin{equation}
  G_{ah} = 0 \,, \qquad G_{hh}=1 \, . 
\end{equation}
$G_{ab}$ is tuned to be
\begin{align}
G_{ab}&=\frac{v^2}{4}\mathcal{F}(h)\delta_{ab} \,.
\end{align}
\subsection{Geometrical form of the $\mathcal{O}(p^2)$ GHEFT Lagrangian}
The lowest order ($\mathcal{O}(p^2)$) GHEFT Lagrangian 
(\ref{eq:ewchlag-with-matter2}) can also be expressed 
in a geometrical form:
\begin{align}
\mathcal{L}
=
\frac{1}{2}g_{ij}(\phi)D_\mu\phi^i D^\mu\phi^j
-V(\phi)
-\dfrac{1}{4} W^a_{\mu\nu} W^{a\mu\nu}
-\dfrac{1}{4} B_{\mu\nu} B^{\mu\nu}
\, ,
\label{eq:Lagrangian}
\end{align}
where $\phi^i$ stands a scalar field multiplet
containing both Higgs bosons $\phi^I$ and the NG bosons $\pi^a$ as its component, 
{\em i.e.,}
\begin{equation}
  \{\phi^i\}  = \{ \pi^a \, , \phi^I \} \, .
\label{eq:gheft-coordinate}
\end{equation}
The geometrical form of the GHEFT Lagrangian (\ref{eq:Lagrangian})
can be understood as a gauged nonlinear sigma model on a scalar
manifold.
The scalar manifold (internal space) is coordinated by the
scalar multiplet $\phi^i$.
Both the metric $g_{ij}(\phi)$ and the potential $V(\phi)$ are
functions of $\phi^i$.
They should be invariant under the $SU(2)_W \times U(1)_Y$ transformation:
\begin{align}
  0 &= w_a^k g_{ij,k} + (w_a^k)_{,i} g_{kj} + (w_a^k)_{,j} g_{ik} \, ,
\label{eq:su2sym-metric}
  \\
  0 &= y^k g_{ij,k} + (y^k)_{,i} g_{kj} + (y^k)_{,j} g_{ik} \, ,
\label{eq:u1sym-metric}
  \\
  0 &= w_a^k V_{,k} \, ,
\label{eq:su2sym-pot}
  \\
  0 &= y^k V_{,k} \, ,
\label{eq:u1sym-pot}
\end{align}
with
\begin{equation}
  g_{ij,k} := \dfrac{\partial}{\partial \phi^k} g_{ij} \, , \qquad
  V_{,k} := \dfrac{\partial}{\partial \phi^k} V \, , \qquad
  (w^k_a)_{,i} := \dfrac{\partial}{\partial \phi^i} w^k_a \, , \qquad
  (y^k)_{,i} := \dfrac{\partial}{\partial \phi^i} y^k \, .
\end{equation}
The $SU(2)_W$ and $U(1)_Y$ Killing vectors are denoted by 
$w_a^k$ ($a=1,2,3$) and $y^k$, respectively, in 
Eqs.~(\ref{eq:su2sym-metric})--(\ref{eq:u1sym-pot}).
The GHEFT Lagrangian (\ref{eq:ewchlag-with-matter2}) 
provides the most general and systematic method to construct
the geometrical form of the Lagrangian (\ref{eq:Lagrangian})
having these symmetry properties 
(\ref{eq:su2sym-metric})--(\ref{eq:u1sym-pot}).
The translation dictionary from the GHEFT Lagrangian (\ref{eq:ewchlag-with-matter2}) to the geometrical form (\ref{eq:Lagrangian}) is given in appendix \ref{app:dictionary}.

The $SU(2)_W \times U(1)_Y$ gauge interactions are introduced
in the scalar sector through the covariant derivative
\begin{align}
D_\mu\phi^i
=
\partial_\mu\phi^i
+
g_W^{} W^a_\mu w^i_a(\phi)
+
g_Y^{} B_\mu y^i(\phi)\, .
\label{eq:covderiv}
\end{align} 
It should be noted that the gauge fields interact with the scalar
sector through the $SU(2)_W \times U(1)_Y$ Killing vectors $w_a^i$ and
$y^i$.

The scalar potential $V(\phi)$ should be minimized at the 
vacuum,
\begin{align}
\VEV{\phi^i}
=
\bar{\phi^i}\, .
\end{align}  
Note that, since the electroweak symmetry is spontaneously broken
at the vacuum, 
the vacuum $\phi^i = \bar{\phi}^i$ cannot be a fixed point of 
the $SU(2)_W \times U(1)_Y$ transformation, {\em i.e.,}
\begin{equation}
  w_a^i(\bar{\phi}) \ne 0 \, , \qquad
  y^i(\bar{\phi}) \ne 0 \, .
\label{eq:killing-vector-at-origin}
\end{equation}
It should be a fixed point of the $U(1)_{\rm em}$ transformation,
\begin{equation}
  w_3^i(\bar{\phi}) + y^i(\bar{\phi}) = 0,
\label{eq:em-unbroken}
\end{equation}
however.
The electroweak gauge bosons ($W$ and $Z$) acquire their masses
\begin{equation}
  M_W^2 \propto g_W^2 g_{ij}(\bar{\phi}) \, 
    w_{1}^i(\bar{\phi}) \, w_{1}^j(\bar{\phi}) \, ,
  \qquad
  M_Z^2 \propto (g_W^2 +g_Y^2 ) g_{ij}(\bar{\phi}) \, 
    w_{3}^i(\bar{\phi}) \, w_{3}^j(\bar{\phi}) \, .
\end{equation}
The Killing vectors at the vacuum (\ref{eq:killing-vector-at-origin})
therefore play the role of the Higgs vacuum expectation value in the SM.
It should be emphasized that the vanishing scalar vacuum expectation value 
$\bar{\phi}^i = 0$ does not imply the electroweak symmetry recovery
in the GHEFT Lagrangian.
Actually, in the GHEFT coordinate (\ref{eq:gheft-coordinate}),
even though 
the vacuum expectation values of the scalar fields are all vanishing
$\bar{\phi}^i=0$, 
the electroweak symmetry is still spontaneously broken by the
non-vanishing Killing vectors at the vacuum (\ref{eq:killing-vector-at-origin}).

The dynamical excitation fields $\varphi^i$ are obtained after the expansion
around the vacuum,
\begin{align}
\phi^i 
=
\bar{\phi^i}
+
\varphi^i\, .
\end{align} 
The scalar manifold metric $g_{ij}$ is expanded as
\begin{equation}
  g_{ij} = \bar{g}_{ij} + \bar{g}_{ij,k}\, \varphi^k 
          + \dfrac{1}{2} \bar{g}_{ij,kl}\, \varphi^k \varphi^l
          + \cdots\, ,
\end{equation}
with
\begin{equation}
  \bar{g}_{ij} := g_{ij}(\bar{\phi}), \qquad
  \bar{g}_{ij,k} := \left. 
    \dfrac{\partial}{\partial \phi^k} g_{ij}(\phi)
  \right|_{\phi=\bar{\phi}}, \qquad
  \bar{g}_{ij,kl} := \left. 
    \dfrac{\partial}{\partial \phi^l} 
    \dfrac{\partial}{\partial \phi^k} 
    g_{ij}(\phi)
  \right|_{\phi=\bar{\phi}}\, , \qquad
  \cdots .
\end{equation}
In a similar manner, the potential term is expanded as
\begin{equation}
  V(\phi)  = \bar{V} + \bar{V}_{,i} \varphi^i
     + \dfrac{1}{2} \bar{V}_{,ij} \varphi^i \varphi^j
     + \dfrac{1}{3!} \bar{V}_{,ijk} \varphi^i \varphi^j \varphi^k
     + \dfrac{1}{4!} \bar{V}_{,ijkl} \varphi^i \varphi^j \varphi^k \varphi^l
     + \cdots\, \label{eq:spotential} 
,
\end{equation}
with
\begin{equation}
  \bar{V} := V \biggr|_{\phi=\bar{\phi}} \, , \quad
  \bar{V}_{,i} := \dfrac{\partial}{\partial \phi^i} 
                 V \biggr|_{\phi=\bar{\phi}} \, , \quad 
  \bar{V}_{,ij} := \dfrac{\partial}{\partial \phi^j} 
                  \dfrac{\partial}{\partial \phi^i} 
                   V \biggr|_{\phi=\bar{\phi}} \, , \quad 
  \cdots .
\end{equation}
Since the potential $V$ is minimized at the vacuum, 
the potential should satisfy
\begin{equation}
  \bar{V}_{,i} = 0\, .  
\label{eq:V-at-vacuum1}
\end{equation}

The scalar manifold is coordinated by the scalar field multiplet
$\phi^i$.  
Hereafter, we normalize/diagonalize the coordinate $\phi^i$ as
\begin{equation}
  \bar{g}_{ij} = \delta_{ij}\, ,
\label{eq:g-at-vacuum}
\end{equation}
and
\begin{equation}
  \bar{V}_{,ij} = \delta_{ij} m_i^2\, ,
\label{eq:V-at-vacuum2}
\end{equation}
so that the excitation fields $\varphi^i$ are canonically normalized
and diagonalized.

\section{Scalar scattering amplitudes and perturbative unitarity}
\label{sec:Amplitude}
We next consider implications of the perturbative unitarity
in the GHEFT framework.
It is well known that, 
in the effective field theory framework, 
the longitudinally polarized electroweak (EW) gauge boson scattering 
amplitudes grow 
in the high energy and tend to cause violations of the perturbative 
unitarity \cite{Zhang:2003it,Chang:2013aya}.
The effective field theory coupling constants need to be arranged 
to keep the perturbative unitarity in the high energy gauge boson 
scattering amplitudes.

For such a purpose, we use the equivalence theorem between the 
longitudinally polarized gauge 
boson scattering amplitudes and the corresponding would-be NG boson
amplitudes \cite{Cornwall:1974km,Chanowitz:1985hj,Gounaris:1986cr,He:1993yd,He:1993qa}.
The equivalence theorem allows us to estimate the longitudinally polarized
gauge boson high energy scattering amplitudes by using the NG boson
amplitudes in the gaugeless limit {\it i.e.,} $g_W=g_Y=0$ with uncertainty of ${\cal O}(M_W^2/E^2)$.
The computation of the amplitudes is simplified greatly in the
gaugeless limit.

Note that the energy growing behavior in the longitudinal polarized 
gauge bosons amplitudes is exactly canceled
in the SM \cite{Cornwall:1973tb,Cornwall:1974km,Llewellyn Smith:1973ey,Lee:1977eg}.  
The energy growing behavior coming from the EW gauge boson exchange 
and contact interaction diagrams is exactly canceled by the Higgs
exchange diagram in the SM.
The Higgs boson plays an essential 
role to keep the perturbative unitarity in the SM.

On the other hand, it is highly non-trivial whether the 
cancellation of the energy growing terms does work or not in the GHEFT. 
In fact, in order to ensure the cancellation, the coupling strengths 
between the Higgs boson(s) and the EW gauge bosons should satisfy 
special conditions known as the  ``unitarity sum rules'' \cite{Gunion:1990kf,Csaki:2003dt,SekharChivukula:2008mj}. 
The unitarity sum rules provide a guiding principle to investigate 
the extended Higgs scenarios in a model-independent manner. 
Model-independent studies on extended EWSB scenarios have been 
done based on the unitarity 
argument \cite{Gunion:1990kf,Grinstein:2007iv,Nagai:2014cua,Abe:2015jra,Abe:2016fjs}. 

We estimate the amplitudes of EW gauge boson scattering by the NG boson scattering 
with the help of the equivalence theorem. 
In subsequent subsections, we explicitly calculate the on-shell amplitudes among the scalar fields $\varphi^i$ in the gaugeless limit,
and express the unitarity sum rules in terms of the scalar manifold's geometry.

\subsection{Scalar scattering amplitudes}
We consider here an $N$-point on-shell scalar scattering amplitude
at the tree-level,
\begin{equation}
  i{\cal M}(123\cdots N)
  :=
  i{\cal M}(\varphi^{i_1}(p_1),
            \varphi^{i_2}(p_2),
            \varphi^{i_3}(p_3),
            \cdots,
            \varphi^{i_N}(p_N)
           )\, ,
\label{eq:amplitude}
\end{equation}
with $p_n$ and $i_n$ ($n=1,2,\cdots, N$) being outgoing momenta 
and the particle species, respectively.

We define
\begin{equation}
  s_{n_1}^{} := p_{n_1}^2\, , \qquad
  s_{n_1 n_2}^{} := (p_{n_1} + p_{n_2})^2\, , \qquad
  s_{n_1 n_2 n_3}^{} := (p_{n_1} + p_{n_2} + p_{n_3})^2\, , \qquad \cdots .
\end{equation}
External momenta $p_n$ are taken on-shell,
\begin{equation}
  s_n^{}  = m_{i_n}^2\, .
\end{equation}
We note
\begin{equation}
  s_{n_1 n_2 n_3}^{} 
  = s_{n_1 n_2}^{}+s_{n_2 n_3}^{}+s_{n_1 n_3}^{}-m_{i_{n_1}}^2-m_{i_{n_2}}^2-m_{i_{n_3}}^2\, , \qquad \cdots .
\end{equation}
The $N$-point amplitude (\ref{eq:amplitude}) can thus be written as a
function of the scalar particle masses $m_{i}^2$ and 
the generalized Mandelstam variables $s_{n_1 n_2}^{}$.

As we will show explicitly below, the three- and four-point on-shell
scattering amplitudes are described in terms of the geometry of the 
scalar manifold,
\begin{align}
  i {\cal M}(123) 
  &= -i \bar{V}_{; (i_1 i_2 i_3)}\, ,
\label{eq:three-point}
\\
  i {\cal M}(1234) 
  &= i M(1234) + i{\cal M}(125) \, [D(s_{12})]_{i_5 i_6} \, i{\cal M}(346)
\nonumber\\
  & \quad
    + i{\cal M}(135) \, [D(s_{13})]_{i_5 i_6} \, i{\cal M}(246)
    + i{\cal M}(145) \, [D(s_{14})]_{i_5 i_6} \, i{\cal M}(236) \,  ,
\end{align}
with
\begin{align}
  iM(1234) &= -i\bar{V}_{;(i_1 i_2 i_3 i_4)}
               -\frac{i}{3} \left(
                 \bar{R}_{i_1 i_3 i_4 i_2} + \bar{R}_{i_1 i_4 i_3 i_2}
                \right) s_{12}
           \nonumber\\
           & 
               -\frac{i}{3} \left(
                 \bar{R}_{i_1 i_2 i_4 i_3} + \bar{R}_{i_1 i_4 i_2 i_3}
                \right) s_{13}
               -\frac{i}{3} \left(
                 \bar{R}_{i_1 i_2 i_3 i_4} + \bar{R}_{i_1 i_3 i_2 i_4}
                \right) s_{14}\, , 
\label{eq:four-point-final}
\end{align}
and
\begin{equation}
  [D(s)]_{ij} := \dfrac{i}{s-m_i^2} \bar{g}_{ij}  \, .
\end{equation}
Here $\bar{V}_{;(i_1 i_2 i_3)}$, $\bar{V}_{;(i_1 i_2 i_3 i_4)}$
and $\bar{R}_{i_1 i_2 i_3 i_4}$ stand for the totally symmetrized
covariant derivatives of the potential and the
Riemann curvature tensor of the scalar manifold at the vacuum.

Let us start with the three-point scalar scattering amplitude ${\cal M}(123)$.
The interaction vertices relevant for this amplitude are
\begin{equation}
  {\cal L}_3 = \dfrac{1}{2} \bar{g}_{ij,k}\, 
               \varphi^k (\partial_\mu \varphi^i) (\partial^\mu \varphi^j) 
              -\dfrac{1}{3!} \bar{V}_{,ijk}\, \varphi^i \varphi^j \varphi^k\, ,
\label{eq:vertices3}
\end{equation}
at the tree-level.
It is straightforward to evaluate the on-shell three-point amplitude 
\begin{align}
  i {\cal M}(123) 
  &=
  \dfrac{i}{2} (\bar{g}_{i_1 i_2, i_3}^{} +\bar{g}_{i_2 i_1, i_3}^{}) (-p_1 \cdot p_2)
   +\dfrac{i}{2} (\bar{g}_{i_2 i_3, i_1}^{} +\bar{g}_{i_3 i_2, i_1}^{}) (-p_2 \cdot p_3)
  \nonumber\\ 
  &  \quad
 +\dfrac{i}{2} (\bar{g}_{i_3 i_1, i_2}^{} +\bar{g}_{i_1 i_3, i_2}^{}) (-p_3 \cdot p_1)
 - i \bar{V}_{,i_1 i_2 i_3}
  \nonumber\\
  &=
  \dfrac{i}{2} \bar{g}_{i_1 i_2, i_3}^{} \left(m_{i_1}^2+m_{i_2}^2-s_{12}\right) 
   +\dfrac{i}{2} \bar{g}_{i_2 i_3, i_1}^{} \left(m_{i_2}^2+m_{i_3}^2-s_{23}\right)
  \nonumber\\ 
  &  \quad
 +\dfrac{i}{2}  \bar{g}_{i_3 i_1, i_2}^{} \left(m_{i_3}^2+m_{i_1}^2-s_{31}\right) 
 - i \bar{V}_{,i_1 i_2 i_3}\, ,
\label{eq:3pt-amp1}
\end{align}
from the vertices in (\ref{eq:vertices3}).
The conservation of the total momentum
\begin{displaymath}
  p_1^\mu + p_2^\mu + p_3^\mu = 0\, ,  
\end{displaymath}
implies
\begin{displaymath}
  s_{12}^{} = (p_1+p_2)^2 = p_3^2 = m_{i_3}^2\, ,
\end{displaymath}
and similarly
\begin{displaymath}
  s_{23}^{} =  m_{i_1}^2\, , \qquad
  s_{31}^{} = m_{i_2}^2\, .
\end{displaymath}
The on-shell three-point amplitude (\ref{eq:3pt-amp1}) can therefore be
expressed as
\begin{align}
  i {\cal M}(123) 
  &=
  \dfrac{i}{2} \bar{g}_{i_1 i_2, i_3}^{}  \left(m_{i_1}^2+m_{i_2}^2-m_{i_3}^2\right)
   +\dfrac{i}{2} \bar{g}_{i_2 i_3, i_1}^{} \left(m_{i_2}^2+m_{i_3}^2-m_{i_1}^2\right)
  \nonumber\\ 
  &  
   \quad
 +\dfrac{i}{2}  \bar{g}_{i_3 i_1, i_2}^{} \left(m_{i_3}^2+m_{i_1}^2-m_{i_2}^2\right) 
 - i\, \bar{V}_{,i_1 i_2 i_3}
  \nonumber\\
  &= 
  \dfrac{i}{2} m_{i_1}^2 \left( 
    \bar{g}_{i_1 i_2, i_3}^{} + \bar{g}_{i_1 i_3, i_2}^{}
   -\bar{g}_{i_2 i_3, i_1}^{} 
  \right)
  + 
  \dfrac{i}{2} m_{i_2}^2 \left( 
    \bar{g}_{i_2 i_3, i_1}^{} + \bar{g}_{i_2 i_1, i_3}^{}
   -\bar{g}_{i_3 i_1, i_2}^{} 
  \right)
  \nonumber\\ 
  &  
   \quad
  + 
  \dfrac{i}{2} m_{i_3}^2 \left(
     \bar{g}_{i_3 i_1, i_2}^{} + \bar{g}_{i_3 i_2, i_1}^{}
    -\bar{g}_{i_1 i_2, i_3}^{}
  \right)
 - i\, \bar{V}_{,i_1 i_2 i_3}\, .
\label{eq:3point-amp5}
\end{align}
Note that the $m_{i_1}^2$, $m_{i_2}^2$ and $m_{i_3}^2$ are related with 
the second derivative of the potential $V_{,ij}$ by
(\ref{eq:V-at-vacuum2}).
The first derivative of the metric tensor in the interaction vertex
(\ref{eq:vertices3}) is related with the 
the Affine connection $\Gamma^l_{jk}$
\begin{equation}
 g_{il} \Gamma^l_{jk} :=
 \dfrac{1}{2} \left[ g_{ij,k} + g_{ki,j} - g_{jk,i} \right]\, .
\end{equation}
The amplitude (\ref{eq:3point-amp5}) can then be rewritten as
\begin{equation}
  i {\cal M}(123) 
  = 
  i\, \bar{V}_{, i_1 l} \bar{\Gamma}^l_{i_2 i_3}
  + 
  i\, \bar{V}_{, i_2 l} \bar{\Gamma}^l_{i_3 i_1}
  + 
  i\, \bar{V}_{, i_3 l} \bar{\Gamma}^l_{i_1 i_2}
 - i\, \bar{V}_{,i_1 i_2 i_3}\, ,
\label{eq:3point-amp6}
\end{equation}
with $\bar{\Gamma}^l_{jk}$ being the Affine connection at the 
vacuum
\begin{equation}
  \bar{\Gamma}^l_{jk} := \Gamma^l_{jk} \biggr|_{\phi=\bar{\phi}}  \, .
\end{equation}
Our final task is to rewrite the amplitude (\ref{eq:3point-amp6}) in terms
of the covariant derivatives of the potential $V$.
It is straightforward to show
\begin{align}
  V_{;ijk}
  &= V_{,ijk} - (\Gamma^l_{ij})_{,k} V_{,l} - \Gamma^l_{ij} V_{,lk}
             - \Gamma^l_{ik} V_{,lj} - \Gamma^l_{jk} V_{,li}
             + \Gamma^l_{ik} \Gamma^m_{lj} V_{,m}
             + \Gamma^l_{jk} \Gamma^m_{li} V_{,m}\, .
\end{align}
Since the first derivative of the potential vanishes at
the vacuum, we obtain
\begin{equation}
  \bar{V}_{;ijk} 
  = \bar{V}_{,ijk} - \bar{\Gamma}^l_{ij} \bar{V}_{,lk}
             - \bar{\Gamma}^l_{ik} \bar{V}_{,lj} 
             - \bar{\Gamma}^l_{jk} \bar{V}_{,li}\, .
\label{eq:third-derivative-pot}
\end{equation}
Moreover, as we see in (\ref{eq:third-derivative-pot}),
$\bar{V}_{;ijk}$ is symmetric under the
$i\leftrightarrow j$, $i\leftrightarrow k$ and 
$j\leftrightarrow k$ exchanges. 
We therefore obtain
\begin{align}
  \bar{V}_{;(ijk)}
  &:= \dfrac{1}{3!} \left[
      \bar{V}_{;ijk}
     +\bar{V}_{;jki}
     +\bar{V}_{;kij}
     +\bar{V}_{;ikj}
     +\bar{V}_{;kji}
     +\bar{V}_{;jik}
      \right]
\nonumber\\
  &= \bar{V}_{;ijk} \, .
\end{align}
It is now easy to obtain a geometrical formula for the three-point
amplitude
\begin{equation}
  i {\cal M}(123) = -i \bar{V}_{; (i_1 i_2 i_3)}\, .
\end{equation}

We next consider the four-point amplitude
\begin{align}
  i {\cal M}(1234) 
  &= iM_0(1234) 
  \nonumber\\
  & \quad 
    + iM(12[5]) [D(s_{12})]_{i_5 i_6} iM(34[6]) 
  \nonumber\\
  &  \quad 
    + iM(13[5]) [D(s_{13})]_{i_5 i_6} iM(24[6]) 
  \nonumber\\
  &  \quad 
    + iM(14[5]) [D(s_{14})]_{i_5 i_6} iM(23[6])\, ,
\label{eq:four-point}
\end{align}
where the first line comes from the four-point contact interaction 
vertices, while the second, the third and the fourth lines are
from the $i_{5}$-particle exchange diagrams in the $s_{12}$, $s_{13}$ 
and $s_{14}$ channels, respectively.
The three-point amplitude $M(ij[k])$ is for on-shell $i$ and $j$, 
allowing off-shell $k$-particle.

We first study $M(12[5])$,
\begin{equation}
  M(12[5]) = - \bar{V}_{;(125)} 
   + \bar{g}_{i_5 i_5'} \bar{\Gamma}^{i_5'}_{i_1 i_2}  (s_{12} - m_{i_5}^2)\, ,
\end{equation}
which can be related with the on-shell three-point amplitude ${\cal M}(125)$
as
\begin{equation}
  M(12[5]) = {\cal M}(125) 
  + \bar{g}_{i_5 i_5'} \bar{\Gamma}^{i_5'}_{i_1 i_2} (s_{12} - m_{i_5}^2)\, .
\end{equation}
It is easy to rewrite the amplitude (\ref{eq:four-point}) as
\begin{eqnarray}
  i {\cal M}(1234) 
  &=& iM(1234) 
  \nonumber\\
  & & 
    + i{\cal M}(125) [D(s_{12})]_{i_5 i_6} i{\cal M}(346) 
  \nonumber\\
  & & 
    + i{\cal M}(135) [D(s_{13})]_{i_5 i_6} i{\cal M}(246) 
  \nonumber\\
  & & 
    + i{\cal M}(145) [D(s_{14})]_{i_5 i_6} i{\cal M}(236)\, ,
\label{eq:four-point1}
\end{eqnarray}
with $M(1234)$ being 
\begin{eqnarray}
  M(1234) &=& M_0(1234) 
  - \bar{g}_{i_5 i_6} \bar{\Gamma}^{i_5}_{i_1 i_2} \bar{\Gamma}^{i_6}_{i_3 i_4} 
    (s_{12} - m_{i_5}^2)
  \nonumber\\
  & &
  - \bar{g}_{i_5 i_6} \bar{\Gamma}^{i_5}_{i_1 i_3} \bar{\Gamma}^{i_6}_{i_2 i_4} 
    (s_{13} - m_{i_5}^2)
  - \bar{g}_{i_5 i_6} \bar{\Gamma}^{i_5}_{i_1 i_4} \bar{\Gamma}^{i_6}_{i_2 i_3} 
    (s_{14} - m_{i_5}^2)
  \nonumber\\
  & & + \bar{V}_{;(i_1 i_2 i_5)} \bar{\Gamma}^{i_5}_{i_3 i_4}
      + \bar{V}_{;(i_1 i_3 i_5)} \bar{\Gamma}^{i_5}_{i_2 i_4}
      + \bar{V}_{;(i_1 i_4 i_5)} \bar{\Gamma}^{i_5}_{i_2 i_3}
  \nonumber\\
  & & + \bar{V}_{;(i_2 i_3 i_5)} \bar{\Gamma}^{i_5}_{i_1 i_4}
      + \bar{V}_{;(i_2 i_4 i_5)} \bar{\Gamma}^{i_5}_{i_1 i_3}
      + \bar{V}_{;(i_3 i_4 i_5)} \bar{\Gamma}^{i_5}_{i_1 i_2}\, .
\end{eqnarray}

The evaluation of the four-point contact interaction contribution
is a bit tedious but straightforward.
We obtain
\begin{eqnarray}
\lefteqn{
  iM_0(1234) = -i \bar{V}_{,i_1 i_2 i_3 i_4}
}
  \nonumber\\
  & & 
     +\dfrac{i}{2} \biggl[
         -\left(\bar{g}_{i_1 i_2, i_3 i_4} + \bar{g}_{i_3 i_4, i_1 i_2} \right) 
          s_{12}
         -\left(\bar{g}_{i_1 i_3, i_2 i_4} + \bar{g}_{i_2 i_4, i_1 i_3} \right) 
          s_{13}
  \nonumber\\
  & &    \qquad
         -\left(\bar{g}_{i_1 i_4, i_2 i_3} + \bar{g}_{i_2 i_3, i_1 i_4} \right) 
          s_{14}
  \nonumber\\
  & &    \qquad
         +\left(\bar{g}_{i_1 i_2, i_3 i_4} + \bar{g}_{i_1 i_3, i_2 i_4} 
                + \bar{g}_{i_1 i_4, i_2 i_3} 
           \right) m_{i_1}^2
         +\left(\bar{g}_{i_2 i_1, i_3 i_4} + \bar{g}_{i_2 i_3, i_1 i_4} 
                + \bar{g}_{i_2 i_4, i_1 i_3} 
           \right) m_{i_2}^2
  \nonumber\\
  & &    \qquad
         +\left(\bar{g}_{i_3 i_1, i_2 i_4} + \bar{g}_{i_3 i_2, i_1 i_4} 
                + \bar{g}_{i_3 i_4, i_1 i_2} 
           \right) m_{i_3}^2
         +\left(\bar{g}_{i_4 i_1, i_2 i_3} + \bar{g}_{i_4 i_2, i_1 i_3} 
                + \bar{g}_{i_4 i_3, i_1 i_2} 
           \right) m_{i_4}^2 \biggr]\, .
  \nonumber\\
  & & 
\end{eqnarray}
Combining these results, we obtain a geometrical formula for 
the on-shell four-point amplitude 
\begin{align}
  iM(1234) &= -i\bar{V}_{;(i_1 i_2 i_3 i_4)}
               -\frac{i}{3} \left(
                 \bar{R}_{i_1 i_3 i_4 i_2} + \bar{R}_{i_1 i_4 i_3 i_2}
                \right) s_{12}
           \nonumber\\
           & 
               -\frac{i}{3} \left(
                 \bar{R}_{i_1 i_2 i_4 i_3} + \bar{R}_{i_1 i_4 i_2 i_3}
                \right) s_{13}
               -\frac{i}{3} \left(
                 \bar{R}_{i_1 i_2 i_3 i_4} + \bar{R}_{i_1 i_3 i_2 i_4}
                \right) s_{14}\, . 
\label{eq:four-point-final}
\end{align}
We used the on-shell condition
\begin{equation}
  s_{12} + s_{13} + s_{14} = m_{i_1}^2 + m_{i_2}^2 +  m_{i_3}^2 + m_{i_4}^2\, ,
\end{equation}
in the computation above.
Here $\bar{R}_{ijkl}$ and $\bar{V}_{;(ijkl)}$ denote the Riemann
curvature tensor and the totally symmetrized covariant
derivatives of the potential at the vacuum:
\begin{equation}
  \bar{R}_{ijkl} := R_{ijkl} \biggr|_{\phi=\bar{\phi}} \, ,
  \qquad
  \bar{V}_{;(ijkl)} := V_{;(ijkl)} \biggr|_{\phi=\bar{\phi}} \, .
\end{equation}
We here give formulas to compute $\bar{R}_{ijkl}$ and
$\bar{V}_{;(ijkl)}$ from the metric tensor $g_{ij}$ and the 
potential $V$:
\begin{equation}
  \bar{R}_{ijkl} = \dfrac{1}{2} \left(
    \bar{g}_{il,jk}+\bar{g}_{jk,il}-\bar{g}_{ik,jl}-\bar{g}_{jl,ik}
  \right)
  +
  \bar{g}_{mn} \left( \bar{\Gamma}^m_{il} \bar{\Gamma}^n_{jk}
                     -\bar{\Gamma}^m_{ik} \bar{\Gamma}^n_{jl} 
  \right)\, ,
\end{equation}
and
\begin{align}
  \bar{V}_{;(ijkl)}
  &= \bar{V}_{,ijkl}
      - \bar{V}_{,ijm} \bar{\Gamma}^m_{kl}
      - \bar{V}_{,klm} \bar{\Gamma}^m_{ij}
      - \bar{V}_{,ikm} \bar{\Gamma}^m_{jl}
      - \bar{V}_{,jlm} \bar{\Gamma}^m_{ik}
      - \bar{V}_{,ilm} \bar{\Gamma}^m_{jk}
      - \bar{V}_{,jkm} \bar{\Gamma}^m_{il}
  \nonumber\\
  & \quad
   + \bar{V}_{,mn} \left[ 
           \bar{\Gamma}^m_{ij} \bar{\Gamma}^n_{kl}
         + \bar{\Gamma}^m_{ik} \bar{\Gamma}^n_{jl}
         + \bar{\Gamma}^m_{il} \bar{\Gamma}^n_{jk}
      \right]
  \nonumber\\
  & \quad
      + A_{ijkl} + A_{jikl} + A_{kijl} + A_{lijk} \, ,
\end{align}
with
\begin{align}
A_{ijkl}
  &:= \dfrac{1}{6} \bar{V}_{,im} \bar{g}^{mn} \left[ 
       \bar{g}_{jk,nl} + \bar{g}_{kl,nj} + \bar{g}_{jl,nk} 
      -2(\bar{g}_{nj,kl} +\bar{g}_{nk,jl} +\bar{g}_{nl,jk} )
       \right]
  \nonumber\\
  & \quad + \bar{V}_{,im} \left[
        \bar{\Gamma}^m_{jn}\bar{\Gamma}^n_{kl}
       +\bar{\Gamma}^m_{kn}\bar{\Gamma}^n_{jl}
       +\bar{\Gamma}^m_{ln}\bar{\Gamma}^n_{jk}
       \right]
  \nonumber\\
  & \quad + \frac{1}{3} \bar{V}_{,im} \bar{g}^{mp} \left[
         \bar{\Gamma}^q_{pj} \bar{\Gamma}^n_{kl}
       + \bar{\Gamma}^q_{pk} \bar{\Gamma}^n_{jl}
       + \bar{\Gamma}^q_{pl} \bar{\Gamma}^n_{jk}
      \right] \bar{g}_{qn} \, .
\end{align}

\subsection{Perturbative unitarity}

As we have shown in Eq.~(\ref{eq:four-point-final}), 
the scalar four-point amplitude $M(1234)$ contains the energy-squared 
terms proportional to $s_{12}$, $s_{13}$ and $s_{14}$. 
This implies that the perturbative unitarity of the scattering amplitude 
is generally violated at certain high energy scale
in the GHEFT~(\ref{eq:Lagrangian}).
In order to keep the 
perturbative unitarity in the high energy limit, 
the GHEFT Lagrangian should satisfy special conditions known as the
unitarity sum rules \cite{Gunion:1990kf,Csaki:2003dt,SekharChivukula:2008mj}.
We here give a geometrical interpretation for the unitarity sum rules.

Applying the on-shell condition
\begin{align}
s_{12}+s_{13}+s_{14}
=
m^2_{i_1}+m^2_{i_2}+m^2_{i_3}+m^2_{i_4} 
\end{align}
in the four-point amplitude (\ref{eq:four-point-final})
we obtain
\begin{align}
iM(1234)
=&
-\frac{i}{3}(
\bar{R}_{i_1i_3i_4i_2}
+
\bar{R}_{i_1i_4i_3i_2}
-
\bar{R}_{i_1i_2i_3i_4}
-
\bar{R}_{i_1i_3i_2i_4}
)s_{12}\nonumber\\
&-\frac{i}{3}(
\bar{R}_{i_1i_2i_4i_3}
+
\bar{R}_{i_1i_4i_2i_3}
-
\bar{R}_{i_1i_2i_3i_4}
-
\bar{R}_{i_1i_3i_2i_4}
)s_{13}+\mathcal{O}(E^0).
\end{align}
Therefore, the unitarity sum rules can be summarized in the geometrical
language as
\begin{align}
\bar{R}_{i_1i_3i_4i_2}
+
\bar{R}_{i_1i_4i_3i_2}
-
\bar{R}_{i_1i_2i_3i_4}
-
\bar{R}_{i_1i_3i_2i_4}
&=
0
\label{eq:unitarity1} \, ,\\
\bar{R}_{i_1i_2i_4i_3}
+
\bar{R}_{i_1i_4i_2i_3}
-
\bar{R}_{i_1i_2i_3i_4}
-
\bar{R}_{i_1i_3i_2i_4}
&=
0
\label{eq:unitarity2} \, .
\end{align}
Note that the Riemann curvature tensor $R_{ijkl}$ is antisymmetric
under the $k\leftrightarrow l$ exchange:
\begin{equation}
  R_{ijkl} \equiv -R_{ijlk} \, .
\end{equation}
The unitarity sum rules (\ref{eq:unitarity1}) can thus be rewritten
as
\begin{equation}
  2 \bar{R}_{i_1 i_3 i_4 i_2} - \bar{R}_{i_1 i_4 i_2 i_3} -\bar{R}_{i_1 i_2 i_3 i_4} = 0
\, .
\label{eq:unitarity3}
\end{equation}
The Bianchi identity
\begin{equation}
  R_{ijkl} + R_{iklj} + R_{iljk} \equiv 0
\label{eq:r-bianchi}
\end{equation}
can be expressed as
\begin{equation}
  - \bar{R}_{i_1 i_4 i_2 i_3} -\bar{R}_{i_1 i_2 i_3 i_4} \equiv \bar{R}_{i_1 i_3 i_4 i_2}
 \, ,
\end{equation}
which enables us to simplify the unitarity sum rules (\ref{eq:unitarity3})
further. 
We obtain the sum rules (\ref{eq:unitarity1}) can be expressed in a
simple form:
\begin{equation}
  3\bar{R}_{i_1 i_3 i_4 i_2}  = 0 \, .
\end{equation}
The unitarity sum rules (\ref{eq:unitarity1})
and (\ref{eq:unitarity2}) can be expressed in a compact form:
\begin{equation}
  \bar{R}_{ijkl} = 0 \, .
\label{eq:unitarity-geo}  
\end{equation}
Note that the unitarity sum rules (\ref{eq:unitarity-geo}) imply the
flatness of the scalar manifold only at the vacuum.
The unitarity conditions (\ref{eq:unitarity-geo}) is lifted to
\begin{equation}
  R_{ijkl} = 0 \, , 
\end{equation}
{\em i.e.,} the complete flatness of the entire scalar manifold at least
in the vicinity of the vacuum, 
by imposing the
perturbative unitarity in the arbitrary $N$-point amplitudes.
See appendix. \ref{app:npoint} for details.

The perturbative unitarity is violated at the certain high energy scale 
in an extended Higgs scenario with a curved scalar manifold.
For instance, if we consider the HEFT with
$\mathcal{F}(h)=(1+(\kappa_V h)/v)^2$ and take $\kappa_V^{}< 1$, 
the corresponding scalar manifold has non-zero curvature proportional 
to $1-\kappa^2_V$ \cite{Alonso:2015fsp, Alonso:2016oah}. 
The model causes the violation of the perturbative unitarity 
at $\Lambda\sim 4\pi v/(1-\kappa^2_V)^{1/2}$.  
In that case, we need to introduce new particles with mass $m\lesssim \Lambda$ and/or to consider non-perturbative effects for 
ensuring the unitarity in the model.

\section{One-loop divergences in the gaugeless limit}
\label{sec:one-loop-gaugeless}
As we have shown in the previous section, the tree-level perturbative
unitarity requires the GHEFT scalar manifold should be flat at the 
vacuum.
What does this imply at the loop level, then?
Refs.~\cite{Alonso:2015fsp, Alonso:2016oah} investigated
the structure of the one-loop divergences in the nonlinear sigma model
Lagrangian (\ref{eq:Lagrangian}).
They found the logarithmic divergences in the scalar one-loop integral 
are described in the gaugeless limit by
\begin{equation}
  \Delta {\cal L}_{\rm div}^{\varphi {\rm -loop}}
  = \dfrac{1}{(4\pi)^2 \epsilon} \left[
      \dfrac{1}{12} \mbox{tr} (Y_{\mu\nu} Y^{\mu\nu})
     +\dfrac{1}{2} \mbox{tr} (X^2)
    \right] \, .
\label{eq:alonso-divergence}
\end{equation}
Here $\epsilon$ is defined as
\begin{equation}
  \epsilon := 4-D,
 \end{equation}
with $D$ being the spacetime dimension.
$Y_{\mu\nu}$ and $X$ are defined as
\begin{align}
{}  [Y_{\mu\nu}]^i{}_j 
  &= R^i{}_{jkl} (D_\mu \phi)^k (D_\nu\phi)^l 
+ W_{\mu\nu}^a (w^i_a)_{;j} + B_{\mu\nu} (y^i)_{;j} \, ,
\\
{}  [X]^i{}_k
  &= R^i{}_{jkl} (D_\mu \phi)^j (D^\mu\phi)^l + g^{ij} V_{;jk} \, ,
\end{align}
with
\begin{equation}
(w^i_a)_{;j}=\frac{\partial}{\partial{\phi}_j}{w}^i_a+{\Gamma}^i_{lj}{w}^l_a
\, , \qquad
(y^i)_{;j} =\frac{\partial}{\partial{\phi}_j}{y}^i+{\Gamma}^i_{lj}{y}^l \, ,
\end{equation}
and $R^i{}_{jkl}=g^{im}R_{mjkl}$.

Remember that the perturbative unitarity implies the flatness at the
vacuum,
\begin{equation}
  \bar{R}_{ijkl} = 0 \, .
\label{eq:flatness_at_vacuum}
\end{equation}
It is easy to see that the unitarity condition (\ref{eq:flatness_at_vacuum})
is enough to guarantee the absence of the divergences 
in the $(\partial_\mu \phi)^4$ type operators, which affect the scalar 
boson high energy four-point scattering amplitudes.
The flatness of the scalar manifold at the vacuum 
(\ref{eq:flatness_at_vacuum}) 
also automatically
guarantees the absence of the divergences in the anomalous triple gauge
boson operators.
These findings are in accord with the general expectations on 
the connections between perturbative unitarity and the absence of new 
counterterms in the one-loop divergences
and also with the explicit heat kernel computations presented 
in Refs.~\cite{Guo:2015isa,Alonso:2017tdy,Buchalla:2017jlu,Alonso:2015fsp,Alonso:2016oah}.

The divergence structure in the operators proportional to
\begin{equation}
  W^a_{\mu\nu} W^{b\mu\nu} \, , \quad
  W^a_{\mu\nu} B^{\mu\nu} \, , \quad
  B_{\mu\nu} B^{\mu\nu} \, ,
\label{eq:gauge-kinetic-operators}
\end{equation}
is not manifest, however.
Note that the oblique correction parameters $S$ and $U$ \cite{Peskin:1990zt} are 
related with the gauge-kinetic-type operators listed 
in (\ref{eq:gauge-kinetic-operators}).
There is no obvious reason to ensure the absence of the one-loop divergences
in the $S$ and $U$ parameters even in the perturbatively unitary models.

Moreover, the one-loop divergence formula (\ref{eq:alonso-divergence})
does not include quantum corrections arising from the
gauge-boson loop diagrams, which should be evaluated to 
deduce the conclusion on the divergence structure for the oblique correction parameters.

In what follows, we explicitly perform the one-loop calculations for both the scalar and gauge loop diagrams.  Our results are consistent with
those of Refs.~\cite{Buchalla:2017jlu,Alonso:2017tdy}, in which the 
complete one-loop divergence formulas 
including gauge-loops and fermionic loop corrections are obtained.
Picking the UV divergent parts from the one-loop functions, we investigate the relationship between the divergence structure and tree-level perturbative unitarity.

\section{Oblique corrections and finiteness conditions}
\label{sec:oblique}

\subsection{Vacuum polarization functions at one-loop}
\label{sec:vacpol}

The electroweak oblique correction parameters $S$ and $U$ are defined
as
\begin{align}
S&:= 16\pi(\Pi'_{33}(0)-\Pi'_{3Q}(0)) \, ,
\label{eq:finitecond1}\\
U&:= 16\pi(\Pi'_{11}(0)-\Pi'_{33}(0)) \, ,
\label{eq:finitecond2}
\end{align}
with $\Pi'_{A}(0)$ being
\begin{equation}
  \Pi'_A(0) := \dfrac{d}{dp^2} \Pi_A(p^2) \biggr|_{p^2=0} \, .
\end{equation}
Here $\Pi_A(p^2)$ stands for the non-SM contribution to the gauge
boson vacuum polarization function in the $A$-channel.
$\Pi_{11}(p^2)$ and $\Pi_{33}(p^2)$ are charged and neutral weak 
$SU(2)_W$ current correlators at momentum $p$, respectively. 
$\Pi_{3Q}(p^2)$ is the correlator between the neutral weak $SU(2)_W$ current and the electromagnetic current. 
Note that, in the GHEFT, a number of scalar particles other than
the 125GeV Higgs contribute to $\Pi_A(p^2)$ at loop.

The oblique correction parameter $T$ is related with Veltman's 
$\rho$ parameter \cite{Veltman:1977kh},
\begin{equation}
  \alpha T := \rho - 1 \, , 
\qquad
  \rho = \dfrac{\dfrac{v_0^2}{4} + \Pi_{11}(0)}
               {\dfrac{v_{Z0}^2}{4} + \Pi_{33}(0)} \, ,
\end{equation}
with $v_0$ and $v_{Z0}$ being the ``bare'' parameters corresponding
to the charged and neutral would-be NG boson decay constants
$v$ and $v_Z$.
The GHEFT Lagrangian loses its predictability on the $T$-parameter,
if we allow to introduce independent counter terms for $v$ and $v_Z$.

On the other hand, if we assume the counter terms for $v$ and $v_Z$
are related with each other,
\begin{equation}
  v_0^2 = v^2 \, ( 1 + \delta_v ) \, , 
  \qquad
  v_{Z0}^2 = v_Z^2 \, ( 1 + \delta_v ) \, , 
\end{equation}
the $\rho$ is calculated as
\begin{equation}
  \rho = \dfrac{v^2}{v_Z^2} 
         \dfrac{1+\delta_v + \dfrac{4}{v^2} \Pi_{11}(0)}
               {1+\delta_v + \dfrac{4}{v_Z^2} \Pi_{33}(0)}
\end{equation}
and we regain a counter-term independent 
predictability on the $\rho$ parameter
\begin{equation}
  \rho = \dfrac{v^2}{v^2_Z} \left(
            1 + \alpha \tilde{T}
         \right) \, ,
\end{equation}
with
\begin{equation}
  \alpha \tilde{T} 
  := 4 \left(
        \dfrac{1}{v^2} \Pi_{11}(0)
       -\dfrac{1}{v_Z^2} \Pi_{33}(0)
       \right) \, .
\label{eq:finitecond3}
\end{equation}

\begin{figure}[t]
 \begin{center}
  \includegraphics[width=40mm]{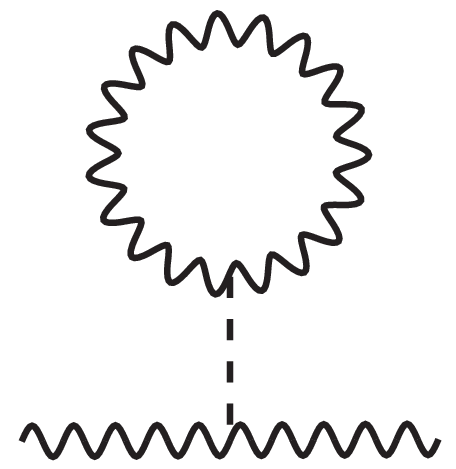}~~~~~~
  \includegraphics[width=40mm]{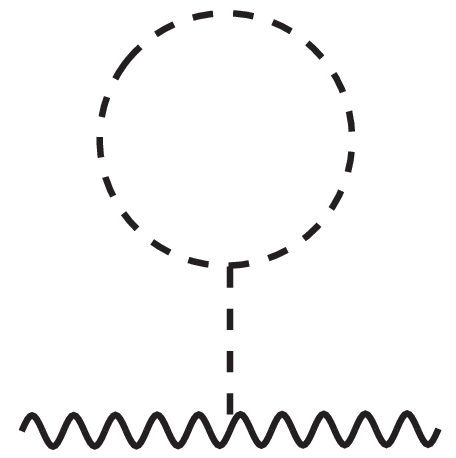}~~~~~~
  \includegraphics[width=40mm]{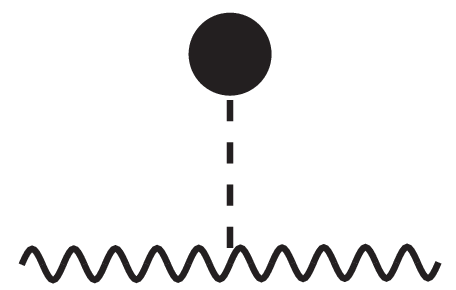}
 \end{center}
 \caption{Feynman diagrams for tadpole contributions
to $\Pi_{11}(0)$ and $\Pi_{33}(0)$ and their counter terms.}
 \label{fig:tadpole}
\end{figure}

In what follows, we calculate $\Pi_{11}$, $\Pi_{33}$, and $\Pi_{3Q}$ 
at one-loop level in the GHEFT and derive the required conditions 
for ensuring the UV finiteness of 
Eqs.~(\ref{eq:finitecond1}), (\ref{eq:finitecond2}) 
and (\ref{eq:finitecond3}).
We apply a background field method \cite{Abbott:1980hw,Honerkamp:1971sh,AlvarezGaume:1981hn,Boulware:1981ns,Howe:1986vm,Fabbrichesi:2010xy} to 
calculate the vacuum polarization functions to keep the 
gauge invariance. 
See Appendix \ref{app:background} for the details of the calculation.
Although there exist UV divergences in $\Pi_{11}(0)$ and $\Pi_{33}(0)$ 
associated with tadpole diagrams as shown in Figure~\ref{fig:tadpole}, 
we assume these UV divergences are canceled by the 
corresponding tadpole counter terms.

\subsubsection{Scalar loop}
\begin{figure}[t]
 \begin{center}
  \includegraphics[width=50mm]{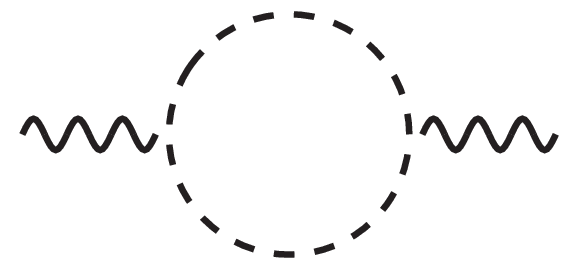}~~~~~~
  \includegraphics[width=50mm]{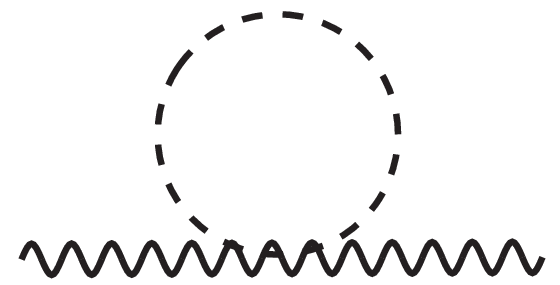}
 \end{center}
 \caption{Feynman diagrams for $\Pi^{\varphi \varphi}_{11}$, $\Pi^{\varphi \varphi}_{33}$, and $\Pi^{\varphi \varphi}_{3Q}$. The internal lines correspond to $\varphi$ fields.}
 \label{fig:xixi}
\end{figure}
Let us start with the scalar loop corrections to the vacuum polarization functions. 
The relevant Feynman diagrams are shown in Figure~\ref{fig:xixi},
which are evaluated to be 
\begin{align}
\Pi^{\varphi \varphi}_{3Q}(p^2)
  &=
   \dfrac{1}{(4\pi)^2} \biggl[
     -2 \sum_{i,j} \, (\bar{w}^i_3)_{;j} \left(
       (\bar{w}^j_3)_{;i}+(\bar{y}^j)_{;i} 
     \right) \, B_{22}(p^2,m^2_i,m^2_j)
\nonumber\\
  &\qquad\qquad
     + \sum_{i,j} \, (\bar{w}^i_3)_{;j} \left(
       (\bar{w}^j_3)_{;i} + (\bar{y}^j)_{;i} 
     \right) \, A(m^2_i) 
     \biggr] \, ,
\end{align}
and
\begin{align}
\Pi^{\varphi \varphi}_{bc}(p^2)
  &=
  \dfrac{1}{(4\pi)^2} \biggl[
    -2 \sum_{i,j} (\bar{w}^i_b)_{;j} \, (\bar{w}^j_c)_{;i} \, 
       B_{22}(p^2,m^2_i,m^2_j)
\nonumber\\
&\qquad\qquad
  + \sum_{i,j}
    \left[ (\bar{w}^i_b)_{;j} \, (\bar{w}^j_c)_{;i}
           + \bar{g}^{ij} \, (\bar{w}^k_b) \, (\bar{w}^l_c) \, \bar{R}_{kilj}
    \right] \, A(m^2_i) 
\biggr] \, ,
\end{align}
for $b,c=1,2,3$.
Here $(\bar{w}^i_{a})_{;j}$ and $(\bar{y}^i)_{;j}$ denote the 
covariant derivatives of the Killing vectors at the vacuum,
\begin{equation}
  (\bar{w}^i_a)_{;j} := (w^i_a)_{;j} \biggr|_{\phi=\bar{\phi}} \, ,
  \qquad
  (\bar{y}^i)_{;j} := (y^i)_{;j} \biggr|_{\phi=\bar{\phi}} \, .
\end{equation}
$A$ and $B_{22}$ are loop functions defined as
\begin{align}
\frac{i}{(4\pi)^2}A(m^2)
&=
\int\frac{d^4 k}{(2\pi)^4}\frac{1}{k^2-m^2},\\
\frac{i}{(4\pi)^2}B_{22}(p^2;m^2_1,m^2_2)
&=
\int\frac{d^4 k}{(2\pi)^4}\frac{k_\mu k_\nu}{(k^2-m^2_1)\left\{(k+p)^2-m^2_2\r\}}\biggl|_{g_{\mu\nu}}.
\end{align}
\subsubsection{Scalar-Gauge loop}
\begin{figure}[t]
 \begin{center}
  \includegraphics[width=50mm]{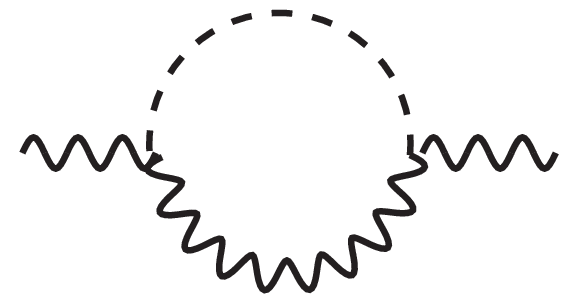}
 \end{center}
 \caption{Feynman diagrams for $\Pi^{\varphi V}_{11}$, $\Pi^{\varphi V}_{33}$, and $\Pi^{\varphi V}_{3Q}$. The internal lines correspond to $\varphi$ and gauge fields.}
 \label{fig:xiV}
\end{figure}
We next calculate the Feynman diagrams shown in Figure~\ref{fig:xiV}. 
In the 't~Hooft-Feynman gauge, we obtain
\begin{align}
\Pi_{3Q}^{\varphi V}(p^2)
  &=0 \, ,
\end{align}
and
\begin{align}
\Pi_{bc}^{\varphi V}(p^2)
  &= -\dfrac{4}{(4\pi)^2} \biggl[
        \sum_{a=1,2} \sum_{i,j} \bar{g}_{ij} \, (G_{Wa})^i_b \, (G_{Wa})^j_c \,
          B_0(p^2, M_W^2, m_i^2) 
\nonumber\\
  & \qquad  +\sum_{i,j} \bar{g}_{ij} \, (G_Z)^i_b \, (G_Z)^j_c \, 
          B_0(p^2, M_Z^2, m_i^2) 
            +\sum_{i,j} \bar{g}_{ij} \, (G_A)^i_b \, (G_A)^j_c \,
          B_0(p^2, 0, m_i^2)
      \biggr] \, ,
\end{align} 
for $b,c=1,2,3$.
Here $(G_{Wa})^i_b$, $(G_{Z})^i_b$, $(G_{A})^i_b$ are defined as
\begin{align}
(G_{Wa})^i_b 
  &:=  g_W \, (\bar{w}_a^i)_{;j} \, \bar{w}_b^j \, , 
  \qquad
  (a=1,2)
\\
(G_{Z})^i_b 
  &:= \dfrac{1}{\sqrt{g_W^2+g_Y^2}} \, \left[
        g_W^2 \, (\bar{w}_3^i)_{;j} - g_Y^2 \, (\bar{y}^i)_{;j}
      \right] \, \bar{w}_b^j \, ,
\\
(G_{A})^i_b
  &:= \dfrac{g_W g_Y}{\sqrt{g_W^2+g_Y^2}} \, \left[
        (\bar{w}_3^i)_{;j} + (\bar{y}^i)_{;j}
      \right] \, \bar{w}_b^j 
\end{align}
and
\begin{equation}
  M_W^2 = \dfrac{g_W^2}{4} v^2 \, , \qquad
  M_Z^2 = \dfrac{g_W^2+g_Y^2}{4} v_Z^2 \, .
\end{equation}
$B_0$ is defined as
\begin{align}
\frac{i}{(4\pi)^2}B_0(p^2;m^2_1,m^2_2)
&=
\int\frac{d^4 k}{(2\pi)^4}\frac{1}{(k^2-m^2_1)\left\{(k+p)^2-m^2_2\r\}}.
\end{align}
\subsubsection{Gauge and Faddeev-Popov (FP) ghost loop}
\begin{figure}[t]
 \begin{center}
  \includegraphics[width=50mm]{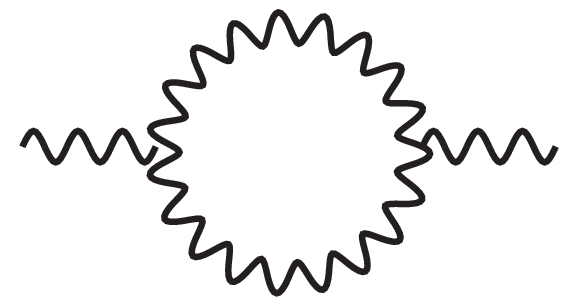}~~~~~~
  \includegraphics[width=50mm]{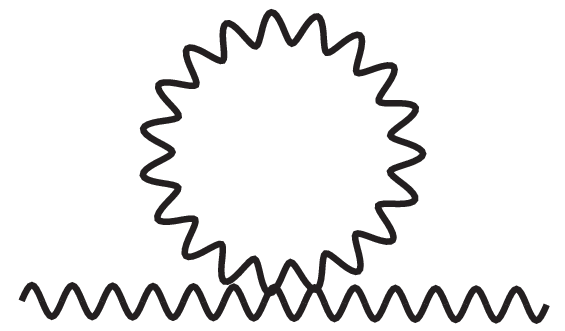}
   \includegraphics[width=50mm]{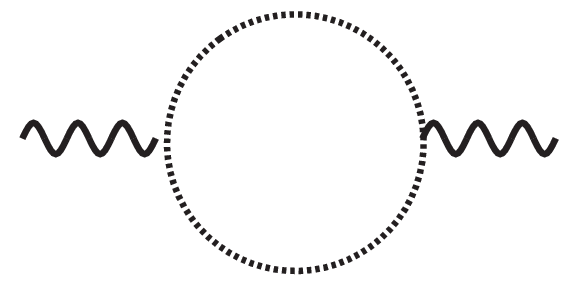}~~~~~~
  \includegraphics[width=50mm]{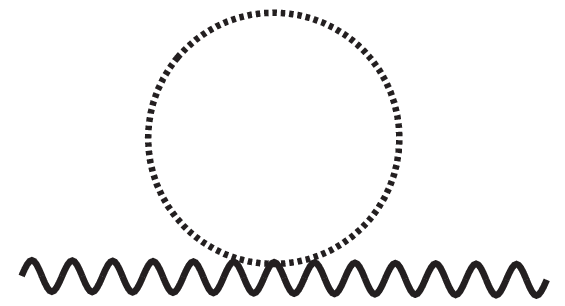}\\
 \end{center}
 \caption{Feynman diagrams for $\Pi^{\rm{Gauge},cc}_{11}$, $\Pi^{\rm{Gauge},cc}_{33}$, and $\Pi^{\rm{Gauge},cc}_{3Q}$. The wavy and dotted lines correspond to gauge fields and Faddeev-Popov ghost fields, respectively.}
 \label{fig:Gaugeloop}
\end{figure}
Finally, we calculate the contributions which are independent of the scalar interactions. The relevant Feynman diagrams are depicted in 
Figure~\ref{fig:Gaugeloop}.
In the 't~Hooft-Feynman gauge, we find the gauge bosons contributions are given by
\begin{align}
\Pi^{\rm{Gauge}}_{11}(p^2)
&= \Pi^{\rm{Gauge}}_{22}(p^2) 
\nonumber\\
&=\frac{4}{(4\pi)^2}\biggl[
-A(M^2_W)-c^2_WA(M^2_Z)-s^2_W A(0)\nonumber\\
&
~~~~~~~~~~~~+2p^2\left(c^2_WB_0(p^2;M^2_Z,M^2_W)+s^2_WB_0(p^2;0,M^2_W)\right)\nonumber\\
&
~~~~~~~~~~~~+4\left(c^2_WB_{22}(p^2;M^2_Z,M^2_W)+s^2_WB_{22}(p^2;0,M^2_W)\right)
\biggr],\\
\Pi^{\rm{Gauge}}_{33}(p^2)
&=
\frac{8}{(4\pi)^2}\biggl[
p^2B_0(p^2;M^2_W,M^2_W)+2B_{22}(p^2;M^2_W,M^2_W)-A(M^2_W)
\biggr]
,\\
\Pi^{\rm{Gauge}}_{3Q}(p^2)
&=
\frac{8}{(4\pi)^2}\biggl[
p^2B_0(p^2;M^2_W,M^2_W)+2B_{22}(p^2;M^2_W,M^2_W)-A(M^2_W)
\biggr],
\end{align}
and Faddeev-Popov (FP) ghost contributions are calculated as
\begin{align}
\Pi^{cc}_{11}(p^2)
&= \Pi^{cc}_{22}(p^2)
\nonumber\\
&=
\frac{2}{(4\pi)^2}\biggl[
A(M^2_W)+c^2_WA(M^2_Z)+s^2_WA(0)\nonumber\\
&
~~~~~~~~-4\left(c^2_WB_{22}(p^2;M^2_Z,M^2_W)+s^2_WB_{22}(p^2;0,M^2_W)\right)
\biggr]
,\\
\Pi^{cc}_{33}(p^2)
&=
-\frac{4}{(4\pi)^2}\biggl[
2B_{22}(p^2;M^2_W,M^2_W)
-A(M^2_W)
\biggr]
,\\
\Pi^{cc}_{3Q}(p^2)
&=
-\frac{4}{(4\pi)^2}\biggl[
2B_{22}(p^2;M^2_W,M^2_W)
-A(M^2_W)
\biggr].
\end{align}
Here $s_W$ and $c_W$ are 
\begin{align}
s_W=\frac{g_Y}{g_Z},\qquad
c_W=\frac{g_W}{g_Z},\qquad
g_Z = \sqrt{g_W^2+g_Y^2} \, .
\end{align}
\subsection{Finiteness of the oblique corrections}
We are now ready to derive the UV finiteness conditions for the
oblique correction parameters at the one-loop level, {\em i.e.,}
the finiteness of 
Eqs.~(\ref{eq:finitecond1}), (\ref{eq:finitecond2}) and
(\ref{eq:finitecond3}).

For the estimation of the UV divergences, 
we regularize the loop functions $A$, $B_0$, and $B_{22}$ by employing the dimensional regularization.  The loop functions are expanded as
\begin{align}
&A(m^2)
=
-\Lambda^2+m^2\ln\frac{\Lambda^2}{\mu^2}-(4\pi)^2A_r(m),
\label{eq:Areg}\\
&B_0(p^2,m^2_1,m^2_2)
=
\ln\frac{\Lambda^2}{\mu^2}+(4\pi)^2B_r(m_1,m_2,p^2),
\label{eq:B0reg}\\
&B_{22}(p^2,m^2_1,m^2_2)
=
-\frac{1}{2}\Lambda^2+\frac{1}{4}\left(m^2_1+m^2_2-\frac{p^2}{3}\right)\ln\frac{\Lambda^2}{\mu^2}+\frac{1}{4}(4\pi)^2B_{0r}(m_1,m_2,p^2),
\label{eq:B22reg}
\end{align}
where the terms proportional to $\Lambda^2$ and $\ln\Lambda^2$ correspond to the terms proportional to $1/(2-D)$ and $1/(4-D)$, respectively. $D$ and $\mu$ denote the spacetime dimension and the renormalization scale, respectively.  
$A_r$, $B_r$, and $B_{0r}$ are $\Lambda$-independent ($\mu$-dependent)
functions. 
The explicit expressions of the $\Lambda$-independent functions are given in Ref. \cite{Nagai:2014cua}.
\subsubsection{$S$ and $U$ parameter}
Let us focus on the UV divergences in 
Eqs.~(\ref{eq:finitecond1}) and (\ref{eq:finitecond2}).
Combining the results derived in subsection \ref{sec:vacpol} and 
Eqs.~(\ref{eq:Areg})-(\ref{eq:B22reg}), 
we find that the UV divergent  parts of $S$ and $U$ are given as
\begin{align}
S_{\rm{div}}
&=
-\dfrac{1}{12\pi}(\bar{w}_3^i)_{;j} (\bar{y}^j)_{;i}\ln\frac{\Lambda^2}{\mu^2}
\label{eq:Slog1},\\
U_{\rm{div}}
&=
\dfrac{1}{12\pi}\left( (\bar{w}_1^i)_{;j}  (\bar{w}_1^j)_{;i}
-(\bar{w}_3^i)_{;j} (\bar{w}_3^j)_{;i} \right)\ln\frac{\Lambda^2}{\mu^2}.
\label{eq:Ulog1}
\end{align}

The gauge boson loops do not contribute to the one-loop divergences in $S$ and $U$ parameters. 
These results are thus identical with the results computed in the 
gaugeless limit \cite{Alonso:2015fsp,Alonso:2016oah}.

\subsubsection{$\Pi_{11}(0)$ and $\Pi_{33}(0)$}

The UV divergences in Eq.~(\ref{eq:finitecond3}) other than the tadpole 
contributions can also be extracted using 
Eqs.~(\ref{eq:Areg})-(\ref{eq:B22reg}).
We obtain
\begin{align}
\left( \dfrac{1}{v^2} \Pi_{11}(0)-\dfrac{1}{v^2_Z}\Pi_{33}(0) \right)_{\rm{div}}
=
\left(\dfrac{1}{v^2} \Pi_{11}(0)-\dfrac{1}{v^2_Z}\Pi_{33}(0)\right)_{\Lambda^2}
+
\left(\dfrac{1}{v^2} \Pi_{11}(0)-\dfrac{1}{v^2_Z}\Pi_{33}(0)\right)_{\ln\Lambda^2},
\label{eq:11-33a}
\end{align}
where
\begin{align}
\biggl( \dfrac{1}{v^2} \Pi_{11}(0)&-\dfrac{1}{v^2_Z}\Pi_{33}(0) \biggr)_{\Lambda^2}
\nonumber\\
=& 
-\dfrac{1}{(4\pi)^2} \left[
   \dfrac{1}{v^2} (\bar{w}_1^i) (\bar{w}_1^j) 
  -\dfrac{1}{v_Z^2} (\bar{w}_3^i) (\bar{w}_3^j) 
 \right] \, \bar{R}_{ikjl} \, \bar{g}^{kl} \, \Lambda^2 \, ,
\label{eq:Tlambda2}
\\
\intertext{and}
\biggl( \dfrac{1}{v^2} \Pi_{11}(0)&-\dfrac{1}{v^2_Z}\Pi_{33}(0) \biggr)_{\ln \Lambda^2}
\nonumber\\
=& \dfrac{1}{(4\pi)^2} \left[
   \dfrac{1}{v^2} (\bar{w}_1^i) (\bar{w}_1^j) 
  -\dfrac{1}{v_Z^2} (\bar{w}_3^i) (\bar{w}_3^j) 
 \right]  
\times
\nonumber\\
& \quad \times 
\biggl\{
  -4g_W^2 (\bar{w}_a^k)_{;i} \, (\bar{w}_a^l)_{;j} \, \bar{g}_{kl}
  -4g_Y^2 (\bar{y}^k)_{;i} \, (\bar{y}^l)_{;j} \, \bar{g}_{kl}
  +\bar{R}_{ikjl} \, (\widetilde{M}^2)^{kl}
  \biggr\} \ln \dfrac{\Lambda^2}{\mu^2} \, ,
\label{eq:Tlnlambda2}
\end{align}
with $(\widetilde{M}^2)^{kl}$ being the scalar boson mass matrix in the
't~Hooft-Feynman gauge:
\begin{equation}
  (\widetilde{M}^2)^{ij}
  := \bar{g}^{ik} \bar{g}^{jl} \bar{V}_{;kl} 
    + g_W^2 (\bar{w}_a^i) (\bar{w}_a^j) 
    + g_Y^2 (\bar{y}^i) (\bar{y}^j) \, ,
\quad
  \bar{V}_{;ij} := V_{;ij} \biggr|_{\phi=\bar{\phi}} \, .
\end{equation}

\section{Perturbative unitarity vs. finiteness conditions}
\label{sec:vs}

We are now ready to discuss the implications of the perturbative
unitarity to the one-loop finiteness of the oblique correction
parameters.
We first concentrate ourselves on the $S$-parameter, the UV-divergence of 
which is given by Eq.~(\ref{eq:Slog1}).
As we stressed in \S.~\ref{sec:one-loop-gaugeless},
since there are no obvious connections between the Riemann
curvature tensor (geometry) $R_{ijkl}$ and the 
$SU(2)_W \times U(1)_Y$ Killing vectors (symmetry) $w_a^i$ and $y^i$, 
the relation between the perturbative unitarity $\bar{R}_{ijkl}=0$
and the one-loop finiteness of the $S$-parameter is not
evidently understood in Eq.~(\ref{eq:Slog1}).

We note, however, that the scalar manifold should be invariant
under the $SU(2)_W \times U(1)_Y$ transformations, and
thus the Killing vectors should satisfy the Killing equations,
\begin{equation}
  0 = (w_a^k) g_{ij,k} + (w_a^k)_{,i} g_{kj} + (w_a^k)_{,j} g_{ik} \, ,
 \quad
  0 = (y^k) g_{ij,k} + (y^k)_{,i} g_{kj} + (y^k)_{,j} g_{ik} \, .
\label{eq:killing}
\end{equation}
There {\em do} exit connections between the geometry ($R_{ijkl}$) and
the symmetry ($w_a^i$ and $y^i$) embedded in the Killing
equations Eqs.~(\ref{eq:killing}).
Moreover, the Killing vectors $w_a^i$ and $y^i$ should obey the
$SU(2)_W \times U(1)_Y$ Lie algebra,
\begin{equation}
  {}[w_a \, , w_b ] = \varepsilon_{abc} w_c \, , 
  \qquad
  {}[w_a \, , y ] = 0 \, ,
\label{eq:lie-algebra}
\end{equation}
with
\begin{equation}
  w_a := w_a^i \dfrac{\partial}{\partial \phi^i} \, , \qquad
  y := y^i \dfrac{\partial}{\partial \phi^i} \, .
\end{equation}

The connections can be studied most easily if we take the Riemann Normal
Coordinate (RNC) around the vacuum $\bar{\phi}$, in which the metric
tensor $g_{ij}(\phi)$ can be expressed in a Taylor-expanded form around
$\bar{\phi}$ as,
\begin{equation}
  g_{ij}(\phi)
  = \delta_{ij} - \dfrac{1}{3} \bar{R}_{ikjl} \varphi^k \varphi^l + \cdots \, ,
\end{equation}
with
\begin{equation}
  \delta_{ij} = \bar{g}_{ij} = g_{ij}(\phi) \biggr|_{\phi=\bar{\phi}} \, ,
  \qquad
  \bar{R}_{ijkl} = R_{ijkl} \biggr|_{\phi=\bar{\phi}} \, .
\end{equation}
Solving the Killing equations (\ref{eq:killing}) in terms of the 
Taylor expansion around the vacuum,
\begin{align}
  w_a^i 
  &= \bar{w}_a^i + (\bar{w}_a^i)_{,j} \varphi^j 
     + \dfrac{1}{2!} (\bar{w}_a^i)_{,jk} \varphi^j \varphi^k
     + \cdots \, ,
  \\
  y^i 
  &= \bar{y}^i + (\bar{y}^i)_{,j} \varphi^j 
     + \dfrac{1}{2!} (\bar{y}^i)_{,jk} \varphi^j \varphi^k
     + \cdots \, ,
\end{align}
we find the Taylor expansion coefficients satisfy
\begin{align}
  0 &= \bar{g}_{ik} (\bar{w}_a^k)_{,j} + \bar{g}_{jk} (\bar{w}_a^k)_{,i} \, ,
    \\
  0 &= \bar{g}_{ik} (\bar{y}^k)_{,j} + \bar{g}_{jk} (\bar{y}^k)_{,i} \, ,
    \\
  (\bar{w}_a^i)_{,jk}
    &= \dfrac{1}{3} \left(
         \bar{R}^i{}_{jkl} + \bar{R}^i{}_{kjl} 
       \right) \bar{w}_a^l \, ,
\label{eq:conn-w}
    \\
  (\bar{y}^i)_{,jk}
    &= \dfrac{1}{3} \left(
         \bar{R}^i{}_{jkl} + \bar{R}^i{}_{kjl} 
       \right) \bar{y}^l \, ,
\label{eq:conn-y}
    \\
     & \quad \vdots
  \nonumber
\end{align}
There certainly exist connections between the geometry $R_{ijkl}$
and the symmetry $w_a^i$ and $y^i$ in  Eqs.~(\ref{eq:conn-w}) and 
(\ref{eq:conn-y}).
However, Eqs.~(\ref{eq:conn-w}) and (\ref{eq:conn-y}) are not enough
to clarify the relation between the perturbative unitarity and 
the $S$-parameter coefficient in (\ref{eq:Slog1}).
Note that the $S$-parameter coefficient is written in terms
of the first derivative of the Killing vectors $(\bar{w}_a^i)_{;j}$
and $(\bar{y}^i)_{;j}$ \, .
We need physical principles to relate $(\bar{w}_a^i)_{;j}$
and $(\bar{y}^i)_{;j}$ with the second derivatives 
$(\bar{w}_a^i)_{,jk}$ and $(\bar{y}^i)_{,jk}$.
Actually, the $SU(2)_W \times U(1)_Y$ Lie algebra (symmetry)
(\ref{eq:lie-algebra}) plays the role.
Plugging Eqs.~(\ref{eq:conn-w}) and (\ref{eq:conn-y}) into 
Eq.~(\ref{eq:lie-algebra}), we obtain
\begin{align}
  (T_a)_j{}^i
  &= \dfrac{1}{2} \varepsilon_{abc} \left([T_b, T_c] \right)_j{}^i
    +\dfrac{1}{2} \varepsilon_{abc} (\bar{w}_b^k) \, (\bar{w}_c^l) \,
     \bar{R}^i{}_{jkl} \, ,
  \\
  0&= \left([T_a, T_Y] \right)_j{}^i + (\bar{w}_a^k) \, (\bar{y}^l) \,
     \bar{R}^i{}_{jkl} \, ,
\end{align}
with $T_a$ and $T_Y$ being matrices denoting the first derivatives
of the $SU(2)_W \times U(1)_Y$ Killing vectors at the vacuum,
\begin{equation}
  (T_a)_j{}^i := (\bar{w}_a^i)_{,j} \, ,
  \qquad
  (T_Y)_j{}^i := (\bar{y}^i)_{,j} \, .
\end{equation}

It is now easy to show
\begin{eqnarray}
  \tr (T_3 T_Y)
  &=& \frac{1}{2} \varepsilon_{3bc} \tr([T_b, T_c] T_Y)
      +\frac{1}{2} \varepsilon_{3bc} (\bar{w}_b^k)\, (\bar{w}_c^l)\,  
\bar{R}^i{}_{jkl} 
      (T_Y)_i{}^j
  \nonumber\\
  &=& \frac{1}{2} \varepsilon_{3bc} \tr([T_c, T_Y] T_b)
      +\frac{1}{2} \varepsilon_{3bc} (\bar{w}_b^k)\,  (\bar{w}_c^l)\,
 \bar{R}^i{}_{jkl} 
      (T_Y)_i{}^j
  \nonumber\\
  &=& -\frac{1}{2} \varepsilon_{3bc} (\bar{w}_c^k)\,  (\bar{y}^l) \,
       \bar{R}^i{}_{jkl} (T_b)_i{}^j
      +\frac{1}{2} \varepsilon_{3bc} (\bar{w}_b^k)\, (\bar{w}_c^l)\, 
 \bar{R}^i{}_{jkl} 
      (T_Y)_i{}^j 
  \nonumber\\
  &=& \frac{1}{2} \varepsilon_{3bc} (\bar{w}_c^k)\, (\bar{w}_3^l)\,
       \bar{R}^i{}_{jkl} (T_b)_i{}^j
      +\frac{1}{2} \varepsilon_{3bc} (\bar{w}_b^k)\, (\bar{w}_c^l)\,
 \bar{R}^i{}_{jkl} 
      (T_Y)_i{}^j\, .
      \label{eq:trace1}
\end{eqnarray}
In the last line of Eq.~(\ref{eq:trace1}), we used the fact that
$U(1)_{\rm em}$ is unbroken at the vacuum Eq.~(\ref{eq:em-unbroken}), {\em i.e.,}
\begin{equation}
  0 = \bar{w}_{3}^i + \bar{y}^i \, .
\end{equation}
Eq.~(\ref{eq:trace1}) can be rewritten in a covariant form
\begin{align}
(\bar{w}^i_{3})_{;j} \, (\bar{y}^j)_{;i}
&=
\dfrac{1}{2}
\left(\varepsilon_{3bc} \, (\bar{w}_c^k)\, (\bar{w}_3^l) \,  
       \bar{R}^i{}_{jkl} \, (\bar{w}^j_{b})_{;i}
      +\varepsilon_{3bc} \, (\bar{w}_b^k)\, (\bar{w}_c^l)\, 
       \bar{R}^i{}_{jkl} \, (\bar{y}^j)_{;i}
\right) \, .
\label{eq:formulaS}
\end{align}
In a similar manner, we obtain the divergent coefficient in the
$U$-parameter (\ref{eq:Ulog1}),
\begin{align}
(\bar{w}^i_{1})_{;j} \, (\bar{w}^j_{1})_{;i}
-(\bar{w}^i_{3})_{;j}\, (\bar{w}^j_{3})_{;i}
&=\frac{1}{2}
\left(
  \varepsilon_{1bc}(\bar{w}^k_b)\, (\bar{w}^l_c) \, 
\bar{R}^i{}_{jkl} \, (\bar{w}^j_{1})_{;i}
-\varepsilon_{3bc}(\bar{w}^k_b)\, (\bar{w}^l_c) \, 
\bar{R}^i{}_{jkl} \, (\bar{w}^j_{3})_{;i}
\right) \, .
\label{eq:formulaU}
\end{align}
Combining Eqs.~(\ref{eq:Slog1}), (\ref{eq:Ulog1}), (\ref{eq:formulaS}), and (\ref{eq:formulaU}), we find 
\begin{align}
S_{\rm{div}}
&=
-\frac{1}{12\pi}
\left(\varepsilon_{3bc} (\bar{w}_c^k) \, (\bar{w}_3^l)\,
       \bar{R}^i{}_{jkl} \, (\bar{w}^j_{b})_{;i}
      +\varepsilon_{3bc} (\bar{w}_b^k)\, (\bar{w}_c^l)\, 
       \bar{R}^i{}_{jkl} \, (\bar{y}^j)_{;i}\right)
      \ln{\dfrac{\Lambda^2}{\mu^2}},
\label{eq:Slog}\\
U_{\rm{div}}
&=
\frac{1}{12\pi}
\left(
\varepsilon_{1bc}(\bar{w}^k_b) \, (\bar{w}^l_c) \, 
\bar{R}^i{}_{jkl} \, (\bar{w}^j_{1})_{;i}
-\varepsilon_{3bc}(\bar{w}^k_b) \, (\bar{w}^l_c) \, 
\bar{R}^i{}_{jkl} \, (\bar{w}^j_{3})_{;i}
\right)
\ln{\dfrac{\Lambda^2}{\mu^2}}.
\label{eq:Ulog}
\end{align} 

The relation between the symmetry and the geometry hidden in
the expressions (\ref{eq:Slog1}) and (\ref{eq:Ulog1}) is now
unveiled in the expressions (\ref{eq:Slog}) and (\ref{eq:Ulog}).
The one-loop divergences of both $S$ and $U$ are proportional to
the Riemann curvature tensor $\bar{R}_{ijkl}$ at the vacuum.
Once the four-point 
tree-level unitarity is ensured, {\em i.e.,} $\bar{R}_{ijkl}=0$, 
then the one-loop finiteness of $S$ and $U$ is automatically 
guaranteed in Eqs.~(\ref{eq:Slog}) and (\ref{eq:Ulog}).
{\footnote{
The relation between the one-loop $S$ divergence and the flatness of the scalar manifold was also pointed out in Ref.~\cite{Alonso:2015fsp} in the context of the HEFT framework, in which 
the connection can be seen more manifestly than in the GHEFT 
framework. 
}}

The physical implications of the $S$ and $U$ parameter formulas 
(\ref{eq:Slog}) and (\ref{eq:Ulog})
can be studied more closely.
Note that both of them vanishes when
\begin{equation}
  (\bar{w}_b^k) \, (\bar{w}_c^l) \, \bar{R}_{ijkl} = 0 \, ,
\label{eq:cond1}
\end{equation}
even if there might exist non-vanishing $\bar{R}_{ijkl}$.
What does the condition (\ref{eq:cond1}) imply, then?
Combining the equivalence theorem and the results presented
in \S.~\ref{sec:Amplitude}, 
we see that the condition (\ref{eq:cond1}) ensures the tree-level unitarity
of the high energy $p$-wave scattering amplitude in the
\begin{equation}
  \pi^b \, \pi^c \to \varphi^i \, \varphi^j
\label{eq:channel1}
\end{equation}
channel.
In the high energy limit, the amplitude (\ref{eq:channel1}) corresponds to the $V^b_L V^c_L \to \varphi^i \, \varphi^j$ scattering amplitudes because of the equivalence theorem. Here $V_L^a$ stands for the longitudinally polarized massive gauge bosons,
$V_L^{1,2}=W_L^{1,2}$ and $V_L^3 = Z_L$.
The one-loop finiteness of the $S$ and $U$ parameters does not require
a completely flat scalar manifold:
the scattering amplitudes $\varphi^i \varphi^j \to \varphi^k \varphi^l$
other than the NG boson channels may still violate the tree-level unitarity.
Once the $p$-wave tree-level unitarity in the channel (\ref{eq:channel1})
is somehow ensured, it is potentially possible
to construct strongly interacting EWSB models without violating the
one-loop finiteness of the $S$ and $U$ parameters.

Moreover, as we see in Appendix~\ref{app:ffamp}, the covariant derivative
of the Killing vector $(w_c^i)_{;j}$ is related with the light-fermion 
scattering amplitudes
\begin{equation}
  f \, \bar{f} \to \varphi^i \, \varphi^j \, .
\end{equation}
Here $f$ (and $\bar{f}$)
stands for light quarks or leptons (light anti-quarks or anti-leptons).
The coefficients in front of the logarithmic divergences in 
Eqs.~(\ref{eq:Slog}) and (\ref{eq:Ulog}) can be expressed in a form
\begin{equation}
  (w_b^k) \, (w_c^l) \, \bar{R}^i{}_{jkl} (\bar{w}_a^j)_{; i} \, .
\end{equation}
This suggests that, assuming negligibly small tree-level 
$\mathcal{O}(p^4)$ contributions,
the precise measurements of the $S$ and $U$ parameters 
can be used to constrain the high energy scattering amplitudes in 
\begin{equation}
  V_L^b \, V_L^c \to \varphi^i \, \varphi^j \, , 
  \qquad
  f\bar{f} \to \varphi^i \, \varphi^j
\label{eq:amps-future}
\end{equation}
channels, which can be tested in future collider experiments.

Finally we make a comment on 
the UV finiteness condition of (\ref{eq:11-33a}).
We find that the UV finiteness of (\ref{eq:11-33a})
is not ensured solely by the flatness of the scalar manifold. 
For an example, even if we assume that the scalar manifold is completely 
flat and $v=v_Z$ at the tree-level, an extra condition
\begin{equation}
  \left[
   (\bar{w}_1^i) (\bar{w}_1^j) 
  -(\bar{w}_3^i) (\bar{w}_3^j) 
 \right] \left[
  g_W^2 (\bar{w}_a^k)_{;i} \, (\bar{w}_a^l)_{;j} \, \bar{g}_{kl}
 +g_Y^2 (\bar{y}^k)_{;i} \, (\bar{y}^l)_{;j} \, \bar{g}_{kl}
\right] = 0 \, .
\label{eq:cond2}
\end{equation}
is required to ensure the finiteness of the one-loop $T$-parameter
correction.
The Georigi-Machacek model 
\cite{Georgi:1985nv,Chanowitz:1985ug,Gunion:1989ci,Gunion:1990dt}
is one of examples where the condition (\ref{eq:cond2}) is not 
satisfied.
We need to introduce independent counter terms for $v$ and $v_Z$ 
in these models.

\section{Summary}
\label{sec:sum}
We have formulated a generalized Higgs effective field theory (GHEFT), 
which includes extra Higgs particles other than the 125\, GeV Higgs boson
as a low energy effective field theory describing the electroweak symmetry
breaking.
The scalar scattering amplitudes are expressed by the geometry 
(Riemann curvature) and the symmetry (Killing vectors) of the scalar 
manifold in the GHEFT.
The one-loop radiative corrections to electroweak oblique corrections 
are also expressed in terms of geometry and symmetry of the scalar manifold.
By using the results,
we have clarified the relationship between the perturbative unitarity and 
the UV finiteness of oblique corrections in the GHEFT.

Especially, we have shown that 
once the tree-level unitarity is ensured, then the $S$ and $U$ parameters'
one-loop finiteness is automatically guaranteed.
The tree-level perturbative unitarity in the scalar amplitudes 
requires the complete flatness of the scalar manifold at the vacuum.
On the other hand, the one-loop finiteness of electroweak oblique 
correction does not require the complete flatness. 
The findings enable us to verify that 
tree-level unitarity condition is stronger than the one-loop UV 
finiteness condition in extended Higgs scenarios.  

We also found connections between the coefficients of $S$ and 
$U$ parameter divergences and the particle scattering amplitudes 
which can be measured in future collider experiments.

We emphasize that future precision measurements of the discovered 
Higgs couplings, cross section, and oblique parameters are quite 
important for investigating the geometry and symmetry of the scalar 
manifold in the generalized Higgs sector.  
Combining collider/precision experimental data with our effective 
theoretical approach, we should be able to obtain new prospects of 
the physics beyond the SM.

\section*{Acknowledgments}
This work was supported by KAKENHI Grant Numbers 16H06490, 18H05542, 19K14701 (R.N.), 16K17697, 18H05543 (K.T.), 15K05047, and 19K03846 (M.T.).

\appendix
\section{A symmetry-geometry dictionary}
\label{app:dictionary}
The metric tensor $g_{ij}(\phi)$ in the geometrical form Lagrangian
(\ref{eq:Lagrangian}) can be computed from the symmetry form Lagrangian
(\ref{eq:ewchlag-with-matter2}).
We obtain
\begin{align}
  g_{11} 
  &= G_{11} 
  - G_{13} \pi^2
  + \dfrac{1}{3} \left(
      -G_{11} \pi^2 \pi^2 + G_{12} \pi^1 \pi^2
    \right)
  + \dfrac{1}{4} G_{33} \pi^2 \pi^2
  +{\cal O}((\pi)^3) \, , 
  \\
  g_{12}
  &= G_{12} 
  + \dfrac{1}{2}\left(
      G_{13} \pi^1 - G_{23} \pi^2
    \right) 
  \nonumber\\
  & \qquad
  + \dfrac{1}{6} \left(
      G_{11} \pi^1 \pi^2 + G_{22} \pi^1 \pi^2 
    - G_{12} \pi^1 \pi^1 - G_{12} \pi^2 \pi^2
    \right)
  - \dfrac{1}{4} G_{33} \pi^1 \pi^2
  +{\cal O}((\pi)^3) \, , 
  \\
  g_{13}
  &= G_{13} - \dfrac{1}{2} G_{33} \pi^2 
    +\dfrac{1}{6} \left(
       - G_{13} \pi^2 \pi^2 + G_{23} \pi^1 \pi^2
     \right)
    - G_{1I} [iQ_\phi]^I{}_J \phi^J \left( 1 - \dfrac{1}{6} \pi^2 \pi^2 \right)
  \nonumber\\
  & \quad
    + \dfrac{1}{2} G_{3I} [iQ_\phi]^I{}_J \phi^J \pi^2
    - \dfrac{1}{6} G_{2I} [iQ_\phi]^I{}_J \phi^J \pi^1 \pi^2
    + {\cal O}((\pi)^3) \, , 
  \\
  g_{22}
  &= G_{22} + G_{23} \pi^1 
   + \dfrac{1}{3} \left(
      - G_{22} \pi^1 \pi^1 + G_{12} \pi^1 \pi^2
     \right) 
   +\dfrac{1}{4} G_{33} \pi^1 \pi^1
    +{\cal O}((\pi)^3) \, , 
  \\
  g_{23}
  &= G_{23} + \dfrac{1}{2} G_{33} \pi^1
    + \dfrac{1}{6} \left(
        G_{13} \pi^1 \pi^2 - G_{23} \pi^1 \pi^1
      \right)
    - G_{2I} [iQ_\phi]^I{}_J \phi^J \left( 1 - \dfrac{1}{6} \pi^1 \pi^1 \right)
  \nonumber\\
  & \quad
    - \dfrac{1}{2} G_{3I} [iQ_\phi]^I{}_J \phi^J \pi^1
    - \dfrac{1}{6} G_{1I} [iQ_\phi]^I{}_J \phi^J \pi^1 \pi^2
    + {\cal O}((\pi)^3) \, , 
  \\
  g_{33}
  &= G_{33} - 2 G_{3I} [iQ_\phi]^I{}_J \phi^J 
   +G_{IJ} [iQ_\phi]^I{}_K [iQ_\phi]^J{}_L \phi^K \phi^L \, , 
  \\
  g_{1I} 
  &= G_{1I} 
    + \dfrac{1}{2} G_{3I} \pi^2
    -\dfrac{1}{6} G_{1I} \pi^2 \pi^2 
    +\dfrac{1}{6} G_{2I} \pi^1 \pi^2 
    +{\cal O}((\pi)^3) \, , 
  \\
  g_{2I}
  &= G_{2I}  
    + \dfrac{1}{2} G_{3I} \pi^1
    +\dfrac{1}{6} G_{1I} \pi^1 \pi^2 
    -\dfrac{1}{6} G_{2I} \pi^1 \pi^1 
    +{\cal O}((\pi)^3) \, , 
  \\
  g_{3I}
  &= G_{3I} - G_{IJ} [iQ_\phi]^J{}_K \phi^K \, , 
  \\
  g_{IJ}
  &= G_{IJ} \, .
\end{align}
Note that the scalar multiplet $\phi^i$ in the geometrical form Lagrangian
(\ref{eq:Lagrangian}) contains both the NGB bosons $\pi^1, \pi^2, \pi^3$
and the Higgs bosons $\phi^I$ as its component, {\em i.e.},
\begin{equation}
  \{ \phi^i \}
  = \{ \pi^1, \pi^2, \pi^3, \phi^I \} \, .
\end{equation}

The $SU(2)_W \times U(1)_Y$ Killing vectors $w^i_a$ and $y^i$ 
are introduced through the covariant derivative (\ref{eq:covderiv}) 
in the geometrical form Lagrangian (\ref{eq:Lagrangian}).
These Killing vectors can be determined from the
infinitesimal $SU(2)_W \times U(1)_Y$ transformation 
properties (\ref{eq:transf1}), (\ref{eq:transf2}) and (\ref{eq:transf3}).
They are
\begin{align}
  (w_1)^1 &= 1 - \dfrac{1}{3} \pi^2 \pi^2 + {\cal O}((\pi)^4) \, , 
  \\
  (w_1)^2 &=     \dfrac{1}{3} \pi^1 \pi^2 + {\cal O}((\pi)^4) \, , 
  \\
  (w_1)^3 &=     -\dfrac{1}{2} \pi^2 
                -\dfrac{1}{24}(\pi^1 \pi^1 +\pi^2 \pi^2) \pi^2
                 +{\cal O}((\pi)^4) \, , 
\\
  (w_1)^I &= -\dfrac{1}{2} \pi^2 [iQ_\phi]^I{}_J \phi^J
             -\dfrac{1}{24} (\pi^1\pi^1 +\pi^2\pi^2) \pi^2
              [iQ_\phi]^I{}_J \phi^J
              +{\cal O}((\pi)^4) \, , 
  \\
  (w_2)^1 &=     \dfrac{1}{3} \pi^1 \pi^2 + {\cal O}((\pi)^4) \, , 
  \\
  (w_2)^2 &= 1 - \dfrac{1}{3} \pi^1 \pi^1 + {\cal O}((\pi)^4) \, , 
  \\
  (w_2)^3 &=     \dfrac{1}{2} \pi^1 
                +\dfrac{1}{24}(\pi^1 \pi^1 +\pi^2 \pi^2) \pi^1
                 +{\cal O}((\pi)^4) \, , 
  \\
  (w_2)^I &= \dfrac{1}{2} \pi^1 [iQ_\phi]^I{}_J \phi^J
            +\dfrac{1}{24} (\pi^1\pi^1 +\pi^2\pi^2) \pi^1
              [iQ_\phi]^I{}_J \phi^J
              +{\cal O}((\pi)^4) \, , 
  \\
  (w_3)^1 &= \pi^2 \, ,
\\
  (w_3)^2 &= -\pi^1 \, ,
\\
  (w_3)^3 &= 1 \, ,
\\
  (w_3)^I &= [iQ_\phi]^I{}_J \phi^J \, , 
\\
  (y)^1 &= 0 \,  ,
\\
  (y)^2 &= 0 \, ,
\\
  (y)^3 &= -1 \, ,
\\
  (y)^I &= 0 \, .
\end{align}

\section{$N$-point amplitude}
\label{app:npoint}
%

Let us consider the Taylor expansion of the scalar manifold metric
tensor $g_{ij}(\phi)$ around the vacuum point $\bar{\phi}^i$,
\begin{equation}
  g_{ij}(\phi)
  = \bar{g}_{ij} 
   + \bar{G}_{ijk} \, \varphi^k
   + \dfrac{1}{2} \bar{G}_{ijkl} \, \varphi^k \varphi^l
   + \dfrac{1}{3!} 
     \bar{G}_{ijklm} \, \varphi^k \varphi^l \varphi^m
   + \dfrac{1}{4!} 
     \bar{G}_{ijklmn} \, \varphi^k \varphi^l \varphi^m \varphi^n
   + \cdots \, ,
\end{equation}
with
$\phi^i = \bar{\phi}^i + \varphi^i$.
The Taylor coefficients can be expressed in terms of the covariant
derivatives of the Riemann curvature tensor in RNC.
They are~\cite{Muller:1997zk,Hatzinikitas:2000xe}
\begin{align}
  \bar{g}_{ij} 
  &= \delta_{ij} \, ,
  \\
  \bar{G}_{ijk} 
  &= 0 \, ,
  \\
  \bar{G}_{ijkl}
  &= \dfrac{2}{3} \bar{R}_{iklj} \, ,
  \\
  \bar{G}_{ijklm}
  &= \bar{R}_{iklj;m} \, ,
  \\
  \bar{G}_{ijklmn}
  &= \dfrac{6}{5} \bar{R}_{iklj;mn} 
    +\dfrac{16}{15} \bar{R}_{iklo} \bar{R}^o{}_{mnj} \, 
\label{eq:six-point-G}
  \\
  & \quad \vdots
  \nonumber
\end{align}
with
\begin{equation}
  \bar{R}_{ijkl} := R_{ijkl} \biggr|_{\phi=\bar{\phi}} \, ,
  \quad
  \bar{R}_{ijkl;m} := R_{ijkl;m} \biggr|_{\phi=\bar{\phi}} \, ,
  \quad
  \bar{R}_{ijkl;mn} := R_{ijkl;mn} \biggr|_{\phi=\bar{\phi}} \, ,
  \quad
  \cdots \, .
\end{equation}
The one-particle-irreducible on-shell $N$-point amplitude 
$M(12\cdots N)$ can thus be expressed\footnote{
Eq.~(\ref{eq:N-point-amp}) can be regarded as a geometrical manifestation
of Weinberg's soft-theorem in on-shell amplitudes.
See Ref.~\cite{Cheung:2017pzi}, for an exampe, for a recent review on
the computational techniques of various on-shell amplitudes including 
nonlinear sigma models.
} as
\begin{equation}
  iM(12\cdots N)
  = -\dfrac{i}{2} \sum_{m<n} s_{mn} 
     \bar{G}_{(i_m i_n)(i_1 i_2 \cdots \check{i}_m \cdots \check{i}_n \cdots i_N)} \, ,
\label{eq:N-point-amp}
\end{equation}
in the gaugeless flat-potential ($V=0$) scalar model.
Scalar particles are all massless in this model.
The indices inside parentheses are understood to be totally symmetrized.
The check symbols on top of $\check{i}_m$ and $\check{i}_n$ 
in the sequence $i_1 i_2 \cdots \check{i}_m \cdots \check{i}_n \cdots i_N$ 
denote the absence of the corresponding indices, {\em i.e.,}
\begin{equation}
  i_1 i_2 \cdots \check{i}_m \cdots \check{i}_n \cdots i_N =
  i_1 i_2 \cdots i_{m-1} \, i_{m+1} \cdots i_{n-1} \, i_{n+1} \cdots i_N \, .
\end{equation}

We show, in this appendix, that the perturbative unitarity
up to the $N$-point amplitudes requires
\begin{equation}
  \bar{R}_{i_1 i_2 i_3 i_4}=0 \, ,
  \quad
  \bar{R}_{i_1 i_2 i_3 i_4; i_5}=0 \, ,
  \quad
  \bar{R}_{i_1 i_2 i_3 i_4; i_5 i_6}=0 \, ,
  \quad
  \cdots
  \quad
  \bar{R}_{i_1 i_2 i_3 i_4; i_5 \cdots i_N}=0 \, .
\label{eq:unitarity-N} 
\end{equation}
The scalar manifold needs to be completely flat at least in the
vicinity of the vacuum.
It should be stressed here, even though we already have a compact 
expression for the $N$-point amplitude (\ref{eq:N-point-amp}), 
it is nontrivial to obtain the unitarity condition 
(\ref{eq:unitarity-N}), since the generalized Mandelstam variables
$s_{mn}$ need to satisfy the momentum conservation conditions
\begin{equation}
  \sum_{n=1}^N s_{mn} = 0 \, ,
\label{eq:momentum-conservation}
\end{equation}
and the conditions coming from the four-dimensional space-time
(Gram determinant conditions)~\cite{Asribekov:1962tgp}.
We need to make full use of the Riemann tensor symmetry in order
to deduce our conclusions (\ref{eq:unitarity-N}).

\noindent\underline{$N=4$} \\
Let us start with the analysis on the four-point scattering amplitude.
We compute the amplitude in the limit
\begin{equation}
  s:= s_{12} = s_{34} = -s_{13} = -s_{24} \ne 0 \, ,
  \quad
  s_{14} = s_{23} = 0 \, .
\label{eq:mandelstam4}
\end{equation}
Clearly the momentum conservation conditions 
(\ref{eq:momentum-conservation}) are satisfied in (\ref{eq:mandelstam4}).
The Gram determinant conditions do not give extra conditions in $N=4$.

In the limit above, the four-point on-shell amplitude behaves as
\begin{equation}
  M(1234) \propto s \, A(1234) \, ,
\label{eq:4-point-amp}
\end{equation}
with
\begin{equation}
  A(1234) := \{(12)|(34)\} + \{(34)|(12)\} - \{(13)|(24)\} - \{(24)|(13)\} \, .
\end{equation}
Here we introduce an abbreviation for the Riemann curvature tensor
\begin{equation}
  \{12|34\} := \bar{R}_{i_1 i_3 i_4 i_2} \, .
\end{equation}
The indices inside parentheses are, again, understood to be 
totally symmetrized.

Considering the amplitude (\ref{eq:4-point-amp}) for large $s$,
we see the perturbative unitarity requires
\begin{equation}
  A(1234) = 0 \, .  
\end{equation}
Using the Riemann curvature tesor symmetry
\begin{equation}
  \{12|34\} = -\{32|14\} = -\{14|32\} = \{34|12\} = \{21|43\} \, ,
\end{equation}
and the first Bianchi identity
\begin{equation}
  \{12|34\} + \{13|42\} + \{14|23\} = 0 \, ,
\end{equation}
the coefficient $A(1234)$ can be computed as
\begin{align}
  A(1234)
  &= 2\{(12)|(34)\} - 2\{(13)|(24)\}
  \nonumber\\
  &= \{12|34\} +\{12|43\} -\{13|24\} -\{13|42\}
  \nonumber\\
  &= \{12|34\} -\{13|42\} +\{14|23\} -\{13|42\}
  \nonumber\\
  &= -3\{13|42\} \, .
\label{eq:1234}
\end{align}
It is now easy to see that the perturbative unitarity
requires the vanishing Riemann curvature tensor at the vacuum,
\begin{equation}
  \bar{R}_{i_1 i_4 i_2 i_3}  = 0 \, .
\label{eq:results-4}
\end{equation}
Taking the external lines $i_1, \cdots , i_4$ arbitrary, 
the result (\ref{eq:results-4}) requires $\bar{R}_{ijkl}=0$, 
which is enough to guarantee the perturbative unitarity
in the arbitrary four-point amplitudes given in the form of
Eq.~(\ref{eq:N-point-amp}).
The considerations in the limit (\ref{eq:mandelstam4}) thus
provide necessary and sufficient conditions for the perturbative
unitarity in the four-point amplitudes.

\noindent\underline{$N=5$} \\
We next consider the five-point scattering amplitude.
Again, we consider the amplitude in the limit
\begin{equation}
  s:= s_{12} = s_{34} = -s_{13} = -s_{24} \ne 0 \, ,
  \quad
  s_{14} = s_{23} = s_{15} = s_{25} = s_{35} = s_{45} = 0 \, .
\label{eq:mandelstam5}
\end{equation}
Note that the fifth particle is considered to be very soft.

We introduce an abbreviation for the covariant derivative of the
Riemann curvature tensor,
\begin{equation}
  \{12|34;5\} := \bar{R}_{i_1 i_3 i_4 i_2 ; i_5} \, .
\end{equation}
The five-point amplitude in the limit behaves as
\begin{equation}
  M(12345) \propto s A(12345) \, ,
\end{equation}
with
\begin{equation}
  A(12345)
  := \{(12)|(34;5)\} + \{(34)|(12;5)\} - \{(13)|(24;5)\} - \{(24)|(13;5)\} \, .
\end{equation}
Using
\begin{equation}
  \{(12)|(34;5)\} = \frac{1}{3} \{(12)|(34);5\}
                 +\frac{1}{3} \{(12)|(35);4\}
                 +\frac{1}{3} \{(12)|(45);3\} \, ,
\end{equation}
we obtain
\begin{align}
  A(12345) 
  &= \dfrac{1}{3} \biggl\{
        \{(12)|(34);5\} + \{(34)|(12);5\}
       -\{(13)|(24);5\} - \{(24)|(13);5\}
     \biggr\}
  \nonumber\\
  &\quad
    + \dfrac{1}{3} \biggl\{
        \{(43)|(25);1\} - \{(42)|(35);1\}
     \biggr\}
    + \dfrac{1}{3} \biggl\{
        \{(34)|(15);2\} - \{(31)|(45);2\}
     \biggr\}
  \nonumber\\
  &\quad
    + \dfrac{1}{3} \biggl\{
        \{(21)|(45);3\} - \{(24)|(15);3\}
     \biggr\}
    + \dfrac{1}{3} \biggl\{
        \{(12)|(35);4\} - \{(13)|(25);4\}
     \biggr\} \, .
\label{eq:12345-1}
\end{align}
The first line in (\ref{eq:12345-1}) can be computed easily
using the result on the four-point amplitude.
The second and the third lines can also be computed in a manner 
similar to Eq.~(\ref{eq:1234}). We find
\begin{equation}
  A(12345)
  = - \{13|42;5\}
    -\frac{1}{2} \{42|53;1\}
    -\frac{1}{2} \{31|54;2\}
    -\frac{1}{2} \{24|51;3\}
    -\frac{1}{2} \{13|52;4\} \, .
\label{eq:12345-2}
\end{equation}
Eq.~(\ref{eq:12345-2}) can be simplifed further with the help of the
second Bianchi identity
\begin{equation}
  \{12|34;5\} + \{14|35;2\} + \{15|32;4\} = 0 \, .
\end{equation}
We obtain
\begin{align}
  A(12345)
  &=- \{13|42; 5\}
    -\dfrac{1}{2} \biggl\{
        \{24|35;1\} + \{25|31;4\}
     \biggr\}
    -\dfrac{1}{2} \biggl\{
        \{13|45;2\} + \{15|42;3\}
     \biggr\}
  \nonumber\\
  &=- \{13|42; 5\}
    +\dfrac{1}{2} \{21|34;5\} + \dfrac{1}{2} \{12|43;5\}
  \nonumber\\
  &= -2\{13|42;5\} \, .
\end{align}
The perturbative unitarity in the five-point amplitude thus requires
\begin{equation}
  \bar{R}_{i_1 i_3 i_4 i_2 ; i_5} = 0 \, .  
\label{eq:results-5}
\end{equation}
It is easy to see that Eq.~(\ref{eq:results-5}) gives
necessary and sufficient conditions for the perturbative
unitarity in the five-point amplitudes.

\noindent\underline{$N=6$} \\
It is now straightforward to derive the perturbative unitarity conditions
for the six-point amplitude $M(123456)$.  
It will be turned out considerations in the limit
\begin{equation}
  s:= s_{12} = s_{34} = -s_{13} = -s_{24} \ne 0 \, ,
\end{equation}
are enough.
Generalized Mandelstam variables other than
$s_{12}$,$s_{34}$,$s_{13}$ and $s_{24}$ are taken to be zero.
Note that the fifth-particle and the sixth-particle are both 
considered
to be very soft in this limit.  Note also 
this choice of the Mandelstam
variables is consistent with the momentum conservation constraints
and the Gram determinant constraints.

We already know the Riemann curvature tensor $\bar{R}_{ijkl}$ vanishes
at the vacuum thanks to the perturabative unitarity of the four-point 
amplitude.
We therefore concentrate ourselves to the $\bar{R}_{ijkl;mn}$ term
in (\ref{eq:six-point-G}).  
The six-point amplitude coming from the $\bar{R}_{ijkl;mn}$ term
in (\ref{eq:six-point-G}) behaves as
\begin{equation}
  M(123456) \propto s \, A(123456) \, ,
\end{equation}
with
\begin{equation}
  A(123456) := 
    \{(12)|(34;56)\} + \{(34)|(12;56)\} 
  - \{(13)|(24;56)\} - \{(24)|(13;56)\}  \, .
\end{equation}
Here we introduce an abbreviation
\begin{equation}
  \{12|34;56\} := \bar{R}_{i_1 i_3 i_4 i_2; i_5 i_6} \, .
\end{equation}
Using 
\begin{align}
  \{(12)|(34;56)\}
  &= \dfrac{1}{6} \{(12)|(34);(56)\}
   +\dfrac{1}{6} \{(12)|(56);(34)\}
   +\dfrac{1}{6} \{(12)|(35);(46)\}
  \nonumber\\
  & \quad
   +\dfrac{1}{6} \{(12)|(46);(35)\}
   +\dfrac{1}{6} \{(12)|(36);(45)\} 
   +\dfrac{1}{6} \{(12)|(45);(36)\} 
\end{align}
we obtain
\begin{equation}
  A(123456) :=  A_1 + A_2 + A_3 + A_4 \, ,
\end{equation}
with
\begin{align}
  A_1 &= 
    \dfrac{1}{6} \biggl\{
      \{(12)|(34);56\} + \{(34)|(12);56\} 
    - \{(13)|(24);56\} - \{(24)|(13);56\}   
    \biggr\} \, ,
  \\
  A_2 &=
    \dfrac{1}{6} \biggl\{
      \{(12)|(35);46\} + \{(12)|(45);36\} 
    + \{(34)|(15);26\} + \{(34)|(25);16\}   
  \nonumber\\
  & \qquad
    - \{(13)|(25);46\}-\{(13)|(45);26\}
    - \{(24)|(15);36\}-\{(24)|(35);16\}
    \biggr\} \, ,
  \\
  A_3 &=
    \dfrac{1}{6} \biggl\{
      \{(12)|(36);45\} + \{(12)|(46);35\} 
    + \{(34)|(16);25\} + \{(34)|(26);15\}   
  \nonumber\\
  & \qquad
    - \{(13)|(26);45\}-\{(13)|(46);25\}
    - \{(24)|(16);35\}-\{(24)|(36);15\}
    \biggr\} \, ,
  \\
  A_4 &=
    \dfrac{1}{6} \biggl\{
      \{(12)|(56);34\} + \{(34)|(56);12\}
     -\{(13)|(56);24\} - \{(24)|(56);13\}
    \biggr\}  \, .
\end{align}
Here we used the fact that the covariant derivatives are 
commutable, justified by the vanishing curvature tensor 
$\bar{R}_{ijkl}=0$ at the vacuum.
The $A_1$ term can be computed easily by using the result of
$A(1234)$.
The $A_2$ and $A_3$ terms can be computed in a manner similar to
the computations of $A(12345)$.  We obtain
\begin{align}
  A_1 = A_2 = A_3 = -\frac{1}{2} \{13|42;56\} \, .
\end{align}
The $A_4$ term can be computed as
\begin{align}
  A_4 
  &= \dfrac{1}{12} \, \biggl\{
      \{12|56;34\} +\{12|65;34\} 
     +\{34|56;12\} +\{34|65;12\}
  \nonumber\\
  & \qquad 
     -\{13|56;24\} -\{13|65;24\}
     -\{24|56;13\} -\{24|65;13\}
    \biggr\} 
  \nonumber\\
  &= \dfrac{1}{12} \, \biggl\{
      \{12|56;34\} +\{12|65;34\} 
     +\{43|65;12\} +\{43|56;12\}
  \nonumber\\
  & \qquad 
     +\{16|53;24\} +\{15|63;24\}
     +\{45|62;13\} +\{46|52;13\}
    \biggr\} 
  \nonumber\\
  &= \dfrac{1}{12} \, \biggl\{
      \left( \{12|56;34\} +\{16|53;24\} \right)
     +\left( \{12|65;34\} +\{15|63;24\} \right)
  \nonumber\\
  & \qquad 
     +\left( \{43|65;12\} +\{45|62;13\} \right)
     +\left( \{43|56;12\} +\{46|52;13\} \right)
    \biggr\} \, .
  \nonumber
\end{align}
Applying the second Bianchi identity, it can be simplified further
\begin{align}
  A_4 
  &= -\dfrac{1}{12} \, \biggl\{
     \{13|52;64\} +\{13|62;54\} + \{42|63;15\} + \{42|53;16\}
    \biggr\} 
  \nonumber\\
  &= -\dfrac{1}{12} \, \biggl\{
     \{31|25;46\} +\{26|31;45\} + \{24|36;15\} + \{35|24;16\}
    \biggr\} 
  \nonumber\\
  &= -\dfrac{1}{12} \, \biggl\{
     \left( \{31|25;46\} + \{35|24;16\} \right) 
    +\left( \{26|31;45\} + \{24|36;15\} \right)
    \biggr\} 
  \nonumber\\
  &= \dfrac{1}{12} \, \biggl\{
     \{34|21;56\} + \{21|34;65\} 
    \biggr\} 
  \nonumber\\
  &= - \dfrac{1}{6} \{13|42; 56\} \, .
\end{align}
The second Bianchi identity is used in the first- and fourth-lines 
in the above calculation.
Combining these results, we find the six-point amplitude can be
expressed in a simple form,
\begin{equation}
  A(123456) = -\dfrac{5}{3} \{13|42;56\} \, .
\end{equation}
The perturbative unitarity condition in the six-point amplitude
$A(123456)=0$ can now be written in terms of the covariant 
derivative of the Riemann curvature
\begin{equation}
  \bar{R}_{i_1 i_4 i_2 i_3; i_5 i_6} = 0 \, .
\end{equation}

It is straightforward to generalize the calculation presented
above to the perturbative unitarity conditions in the $N$-point
amplitude,
\begin{equation}
  \bar{R}_{i_1 i_4 i_2 i_3; i_5 i_6 \cdots i_N} = 0 \, .
\end{equation}
Since the Taylor expansion coefficients of $R_{ijkl}(\phi)$ are
required to vanish at any order, the $N$-point perturbative 
unitarity requires the Riemann curvature to be
\begin{equation}
  R_{ijkl}(\phi) = 0 \, ,
\label{eq:unitarity-final}
\end{equation}
at least in the vicinity of the vacuum.  There may exist
non-perturbative essential singularity type corrections 
to (\ref{eq:unitarity-final}), though.

\section{Background field method}
\label{app:background}
In this appendix, 
we briefly summarize the interaction terms used in the calculation 
of the vacuum polarization functions in the background field method
at the one-loop level. 
The background field method is reviewed  
in Refs. \cite{Abbott:1980hw,Honerkamp:1971sh,AlvarezGaume:1981hn,Boulware:1981ns,Howe:1986vm,Fabbrichesi:2010xy}.

We start with the lowest order ($\mathcal{O}(p^2)$) gauged nonlinear sigma 
model Lagrangian (\ref{eq:Lagrangian}).
Let us first decompose $\phi^i$, $W^a_\mu$ and $B_\mu$ into 
the background fields and the fluctuation fields as
\begin{align}
&\phi^i
:=
\tilde{\phi}^i+\xi^i-\frac{1}{2}\tilde{\Gamma}^{i}_{~jk}\xi^j\xi^k+\cdots \, ,
\label{eq:exp-phi}\\
&W^a_{\mu}
:=
\tilde{W}^a_{\mu}+\mathcal{W}^a_{\mu} \, ,\\
&B_{\mu}
:=
\tilde{B}_{\mu}+\mathcal{B}_{\mu} \, ,
\end{align}
where $\tilde{\phi}^i$, $\tilde{W}^a_\mu$, and $\tilde{B}_\mu$ are 
the background fields.
The dynamical fluctuation fields are denoted by 
$\xi^i$, $\mathcal{W}^a_\mu$, and $\mathcal{B}_\mu$.
$\tilde{\Gamma}^{i}_{~jk}$ represents the Christoffel symbols 
for the metric $g_{ij}$ at $\phi=\tilde{\phi}$. 
The metric tensor and the Killing vector fields are expanded as
\begin{align}
g_{ij}
&=
\tilde{g}_{ij}
+\frac{1}{3}\tilde{R}_{iklj}\xi^k\xi^l+\cdots,\\
w^i_a
&=
\tilde{w}^i_a
+(\tilde{w}^i_a)_{;j} \xi^j
+\frac{1}{3}\tilde{R}^i_{~klj}\tilde{w}^j_a\xi^k\xi^l+\cdots,\\
y^i
&=
\tilde{y}^i
+(\tilde{y}^i)_{;j} \xi^j
+\frac{1}{3}\tilde{R}^i_{~klj}\tilde{y}^j\xi^k\xi^l+\cdots,
\end{align}
where $\tilde{g}_{ij}$, and $\tilde{R}_{ikjl}$ denote the metric, 
and the Riemann curvature tensor evaluated at $\phi=\tilde{\phi}$. 
$\tilde{w}^i_a$ and $\tilde{y}^i$ are the $SU(2)_W$ and $U(1)_Y$
Killing vectors, while 
$(\tilde{w}^i_a)_{;j}$ and $(\tilde{y}^i)_{;j}$ are the covariant derivatives of the Killing vectors evaluated at $\phi=\tilde{\phi}$. 

The Lagrangian (\ref{eq:Lagrangian}) is expanded as
\begin{align}
\mathcal{L}
=
\mathcal{L}^{(0)}+\mathcal{L}^{(1)}+\mathcal{L}^{(2)}+\cdots,
\end{align} 
where $\mathcal{L}^{(n)}$ is of order $n$ in the fluctuation fields.

The quadratic terms $\mathcal{L}^{(2)}$ is given as
\begin{align}
\mathcal{L}^{(2)}
=&
-\frac{1}{2} \mathcal{W}^a_\mu
\left(
-\tilde{D}^2\delta_{ab}\eta^{\mu\nu}+\tilde{D}^\nu\tilde{D}^\mu\delta_{ab}
-g_W^2\tilde{g}_{ij}\tilde{w}^i_a\tilde{w}^j_b\eta^{\mu\nu}
-g_W\tilde{W}^{c\mu\nu}\varepsilon^{abc}
\right)\mathcal{W}^b_\nu\nonumber\\
&
-\frac{1}{2} \mathcal{B}_\mu
\left(
-\partial^2\eta^{\mu\nu}+\partial^\nu\partial^\mu
-g^2_Y\tilde{g}_{ij}\tilde{y}^i\tilde{y}^j\eta^{\mu\nu}
\right)\mathcal{B}_\nu\nonumber\\
&
+g_W g_Y \mathcal{W}^a_\mu
\left(
\tilde{g}_{ij}\tilde{w}^i_a\tilde{y}^j\eta^{\mu\nu}
\right)\mathcal{B}_\nu\nonumber\\
&
+\frac{1}{2} \xi^i
\left(
-\tilde{D}^2\tilde{g}_{ij}
-\tilde{D}_\mu\tilde{\phi}^k\tilde{D}^\mu\tilde{\phi}^l \tilde{R}_{kilj}
-\tilde{V}_{;ij}
\right)
\xi^j\nonumber\\
&
+2g_W \mathcal{W}^a_\mu
\left(
\tilde{g}_{jk}(\tilde{w}^k_a)_{;i}\tilde{D}^\mu\tilde{\phi}^j
\right)\xi^i\nonumber\\
&
+2g_Y \mathcal{B}_\mu
\left(
\tilde{g}_{jk}(\tilde{y}^k)_{;i}\tilde{D}^\mu\tilde{\phi}^j
\right)\xi^i\nonumber\\
&
-g_W 
\tilde{g}_{ji}\tilde{w}^j_a(\tilde{D}^\mu\mathcal{W}^a_\mu)
\xi^i
-g_Y 
\tilde{g}_{ji}\tilde{y}^j(\partial^\mu\mathcal{B}_\mu)
\xi^i \, ,\label{eq:appeqL}
\end{align}
with $\eta^{\mu\nu}$ being the space-time metric.
Here we define
\begin{align}
&\tilde{D}_\mu\mathcal{W}^a_\mu
:=
\partial_\mu\mathcal{W}^a_\mu
-
g_W\varepsilon^{abc}\tilde{W}^b_\mu \mathcal{W}^c_\mu,\\
&\tilde{D}_\mu\tilde{\phi}^i
:=
\partial_\mu\tilde{\phi}^i
+g_W\tilde{W}^a_\mu\tilde{w}^i_a
+g_Y\tilde{B}_\mu\tilde{y}^i,\\
&\tilde{D}_\mu\xi^i
:=
\partial_\mu\xi^i
+\tilde{\Gamma}^{i}_{~kj}(\partial_\mu\tilde{\phi})^j\xi^k
+g_W\tilde{W}^a_\mu
(\tilde{w}^i_a)_{;j}
\xi^j
+g_Y\tilde{B}_\mu
(\tilde{y}^i)_{;j}
\xi^j,\\
&\tilde{V}_{;ij}
:=
V_{;ij}\biggr|_{\phi=\tilde{\phi}}.
\end{align}

In order to 
compute radiative corrections, we introduce the gauge fixing action,
\begin{align}
\mathcal{L}_{\rm{GF}}
:=
-\frac{1}{2\alpha_W} G^a_WG^a_W
-\frac{1}{2\alpha_Y} G_YG_Y,
\end{align}
where
\begin{align}
G^a_W
&:=\tilde{D}^\mu\mathcal{W}^a_\mu
  -g_W \alpha_W\tilde{g}_{ij}\tilde{w}^i_a\xi^j,\\
G_Y
&:=\partial^\mu\mathcal{B}_\mu
  -g_Y \alpha_Y\tilde{g}_{ij}\tilde{y}^i\xi^j.
\end{align}
with $\alpha_W$ and $\alpha_Y$ being gauge fixing parameters for $SU(2)_W$ and $U(1)_Y$ symmetry, respectively. In the one-loop calculation performed in \S.~\ref{sec:vacpol}, we take $\alpha_W=\alpha_Y=1$ ('t Hooft Feynman gauge). 

We then obtain
\begin{align}
&\mathcal{L}^{(2)}+\mathcal{L}_{\rm{GF}}\nonumber\\
&=
-\frac{1}{2} \mathcal{W}^a_\mu
\left(
-\tilde{D}^2\delta_{ab}\eta^{\mu\nu}+\left(1-\frac{1}{\alpha_W}\right)\tilde{D}^\mu\tilde{D}^\nu\delta_{ab}
-g_W^2\tilde{g}_{ij}\tilde{w}^i_a\tilde{w}^j_b\eta^{\mu\nu}
-2g_W \tilde{W}^{c\mu\nu}\varepsilon^{abc}
\right)\mathcal{W}^b_\nu\nonumber\\
&
-\frac{1}{2} \mathcal{B}_\mu
\left(
-\partial^2\eta^{\mu\nu}+\left(1-\frac{1}{\alpha_Y}\right)\partial^\mu\partial^\nu
-g^2_Y\tilde{g}_{ij}\tilde{y}^i\tilde{y}^j\eta^{\mu\nu}
\right)\mathcal{B}_\nu\nonumber\\
&
+g_W g_Y \mathcal{W}^a_\mu
\left(
\tilde{g}_{ij}\tilde{w}^i_a\tilde{y}^j\eta^{\mu\nu}
\right)\mathcal{B}_\nu\nonumber\\
&
+\frac{1}{2} \xi^i
\left(
-\tilde{D}^2\tilde{g}_{ij}
-\tilde{D}_\mu\tilde{\phi}^k\tilde{D}^\mu\tilde{\phi}^l\tilde{R}_{kilj}
-\alpha_W g_W^2\tilde{g}_{lj}\tilde{g}_{ki}\tilde{w}^l_a\tilde{w}^k_a
-\alpha_Y g^2_Y\tilde{g}_{lj}\tilde{g}_{ki}\tilde{y}^l\tilde{y}^k
-\tilde{V}_{;ij}
\right) \xi^j
\nonumber\\
&
+2g_W \mathcal{W}^a_\mu
\left(
\tilde{g}_{jk}(\tilde{w}^k_a)_{;i}\tilde{D}^\mu\tilde{\phi}^j
\right)\xi^i
+2g_Y \mathcal{B}_\mu
\left(
\tilde{g}_{jk}(\tilde{y}^k)_{;i}\tilde{D}^\mu\tilde{\phi}^j
\right)\xi^i.
\end{align}

We also need to introduce the Faddeev-Popov (FP) action 
\begin{align}
\mathcal{L}_{\rm{FP}}
&:=
ig_W \bar{c}^a_W\frac{\delta G^a_W}{\delta\theta^b_W}c^b_W
+ig_Y \bar{c}_Y\frac{\delta G_Y}{\delta\theta_Y}c_Y
+ig_Y \bar{c}^a_W\frac{\delta G^a_W}{\delta\theta_Y}c_Y
+ig_W \bar{c}_Y\frac{\delta G_Y}{\delta\theta^b_W}c^b_W,
\end{align}
associated with the gauge fixing term where
\begin{align}
\frac{\delta G^a_W}{\delta\theta^c_W}
&:=
-\frac{1}{g_W}\l[
(\partial^\mu\delta^{ad}-g_W \varepsilon^{abd}\tilde{W}^{b\mu})
(\partial_\mu\delta^{dc}-g_W \varepsilon^{dec}\tilde{W}^{e}_{\mu})
+g_W^2\alpha_W\tilde{g}_{ij}\tilde{w}^i_a\tilde{w}^j_c\r]+\mathcal{O}(\xi)\\
\frac{\delta G^a_W}{\delta\theta_Y}
&:=
-g_W\alpha_W\tilde{g}_{ij}\tilde{w}^i_a\tilde{y}^j+\mathcal{O}(\xi),\\
\frac{\delta G_Y}{\delta\theta^b_W}
&:=
-g_Y\alpha_Y\tilde{g}_{ij}\tilde{y}^i\tilde{w}^j_b+\mathcal{O}(\xi),\\
\frac{\delta G_Y}{\delta\theta_Y}
&:=
-\frac{1}{g_Y}\left(\partial^2+g^2_Y\alpha_Y\tilde{g}_{ij}\tilde{y}^i\tilde{y}^j\right)+\mathcal{O}(\xi).
\end{align}
The $\mathcal{L}_{\rm{FP}}$ is expanded as
\begin{align}
\mathcal{L}_{\rm{FP}}
&=
i\left(
(\tilde{D}^\mu\bar{c}^a_W)\tilde{D}_\mu c^a_W-g_W^2\alpha_W\tilde{g}_{ij}\tilde{w}^i_a\tilde{w}^j_b\bar{c}^a_Wc^b_W
\right)\nonumber\\
&
+i\left(
(\partial^\mu\bar{c}_Y)\partial_\mu c_Y-g^2_Y\alpha_Y\tilde{g}_{ij}\tilde{y}^i\tilde{y}^j\bar{c}_Yc_Y
\right)\nonumber\\
&-ig_W g_Y\alpha_W\bar{c}^a_W\tilde{g}_{ij}\tilde{w}^i_a\tilde{y}^j c_Y\nonumber\\
&-ig_W g_Y\alpha_Y\bar{c}_Y\tilde{g}_{ij}\tilde{y}^i\tilde{w}^j_a c^a_W
+\cdots,
\label{eq:LFP}
\end{align}
where
\begin{align}
&\tilde{D}_\mu c^a_W
  :=\partial_\mu c^a_W-g_W \varepsilon^{abc}\tilde{W}^{b}_\mu c^c_W,\\
&\tilde{D}_\mu \bar{c}^a_W
  :=\partial_\mu \bar{c}^a_W-g_W \varepsilon^{abc}\tilde{W}^{b}_\mu\bar{c}^c_W.
\end{align}
In Eq.~(\ref{eq:LFP}), we only show the quadratic terms of the fluctuation fields.

The one-loop vacuum polarizations among the electroweak gauge boson can be evaluated by using the quadratic Lagrangian, $\mathcal{L}^{(2)}+\mathcal{L}_{\rm{GF}}+\mathcal{L}_{\rm{FP}}$. In \S.~\ref{sec:vacpol}, we calculate the one-loop diagrams where the internal lines are the fluctuation fields or FP ghosts.

\section{$f\bar{f} \to \varphi^i \varphi^j$ amplitude}
\label{app:ffamp}
The four-point scalar boson scattering amplitudes are described
by the Riemann curvature tensor $\bar{R}_{ijkl}$ and
the covariant derivatives of the potential 
$\bar{V}_{;ij}$, $\bar{V}_{;ijk}$, $\bar{V}_{;ijkl}$ at the vacuum
in the nonlinear sigma model, as we have shown in \S.~\ref{sec:Amplitude}.
These tensors can, therefore, be measured through the measurements
of the scalar boson scattering cross sections.

When we consider a gauged nonlinear sigma model, the derivative 
$\partial_\mu \phi^i$ is replaced by the covariant one $(D_\mu \phi)^i$
\begin{equation}
  (D_\mu \phi)^i := \partial_\mu \phi^i + g_V V_\mu \, v^i(\phi) \, ,
\label{eq:covariant-derivative}
\end{equation}
with $V_\mu$ and $v^i(\phi)$ being a gauge field and its corresponding
Killing vector.
The gauge coupling strength is denoted by $g_V$ in 
(\ref{eq:covariant-derivative}).
If the Killing vector $v^i(\phi)$ does not vanish at the vacuum
\begin{equation}
  \bar{v}^i := v^i(\phi) \biggr|_{\phi=\bar{\phi}}  \ne 0 \, ,
\end{equation}
it implies that the gauge symmetry is spontaneously broken, and
the gauge boson $V_\mu$ acquires its mass
\begin{equation}
  M_V^2 = g_V^2 \, \bar{g}_{ij} \, (\bar{v}^i) \, (\bar{v}^j) \, ,
\end{equation}
with 
\begin{equation}
  \bar{g}_{ij} := g_{ij}(\phi) \biggr|_{\phi=\bar{\phi}} \, .
\end{equation}
The magnitude of the Killing vector at the vacuum,
$\bar{g}_{ij} \, (\bar{v}^i) \, (\bar{v}^j)$, 
can therefore be determined by the gauge boson mass measurement.

How can we measure the first covariant derivative of the
Killing vector
\begin{equation}
  (\bar{v}^i)_{;j} := (v^i)_{;j} \biggr|_{\phi=\bar{\phi}}
\end{equation}
from experimental observables in the gauged nonlinear sigma model, 
then?
We address the issue in this appendix and show that 
the process $f\bar{f} \to V_\mu \to \varphi^i \, \varphi^j$ can 
be used to determine $(\bar{v}^i)_{;j}$.
Here we introduce a spin-$1/2$ fermion multiplet $f$.
It couples with the gauge field $V_\mu$ through its covariant derivative
\begin{equation}
  D_\mu f := \partial_\mu f + g_V \, V_\mu T_V^{(f)} f \, ,
\end{equation}
with $T_V^{(f)}$ being the charge matrix of the fermion multiplet $f$.
Note that, in order to keep the Lagrangian gauge invariant,
the fermion current 
\begin{equation}
  J_V^\mu := \bar{f} \gamma^\mu T_V^{(f)} f
\end{equation}
must be conserved
\begin{equation}
  0 = \partial_\mu J_V^\mu \, .
\end{equation}

In order to calculate the $f\bar{f} \to V_\mu \to \varphi^i \varphi^j$
amplitude, we consider the gauge interaction Lagrangian
\begin{equation}
  {\cal L}_{V\phi} = g_V \, V_\mu \, g_{ij}(\phi) \, 
                   (\partial^\mu \phi^i) \, 
                   v^j(\phi) \, ,
\label{eq:lagrangian2}
\end{equation}
which can be derived from the nonlinear sigma model kinetic term,
\begin{equation}
  \dfrac{1}{2} g_{ij}(\phi) \, (D_\mu \phi)^i \, (D^\mu \phi)^j
  \in {\cal L} \, .
\end{equation}
Expanding the scalar manifold metric $g_{ij}(\phi)$ and 
the Killing vector $v^j(\phi)$ by the dynamical excitation field
$\varphi^i$, we obtain
\begin{align}
  g_{ij}(\phi) 
  &= \bar{g}_{ij} + \varphi^k \bar{g}_{ij,k} + \cdots \, ,
\\
  v^j(\phi) 
  &= \bar{v}^j + \varphi^k (\bar{v}^j)_{,k} + \cdots \, ,
\end{align}
with
\begin{equation}
  \bar{g}_{ij,k}  := 
  \dfrac{\partial}{\partial \phi^k} g_{ij} \biggr|_{\phi=\bar{\phi}} \, ,
\qquad
  (\bar{v}^j)_{,k} :=
  \dfrac{\partial}{\partial \phi^k} v^j \biggr|_{\phi=\bar{\phi}} \, ,
\end{equation}
and
\begin{equation}
  \phi^i = \bar{\phi}^i + \varphi^i \, .
\end{equation}
The interaction Lagrangian (\ref{eq:lagrangian2}) can be expanded as
\begin{equation}
  {\cal L}_{V\phi} = g_V V_\mu 
    \bar{g}_{ij} \, (\partial^\mu \varphi^i) \, (\bar{v}^j)
   +g_V V_\mu \,  (\partial^\mu \varphi^i) \varphi^k \left(
      \bar{g}_{ij} \, (\bar{v}^j)_{,k} +\bar{g}_{ij,k} \, (\bar{v}^j)
  \right) + \cdots \, .
\label{eq:action2}
\end{equation}
Note that on-shell amplitudes are not affected by total derivative
terms in the Lagrangian.
The interaction Lagrangian (\ref{eq:action2}) can thus be replaced by
\begin{align}
  {\cal L}'_{V\phi}
  &=
  {\cal L}_{V\phi} - \frac{1}{2} \partial^\mu \left(
    g_V V_\mu \varphi^i \varphi^k (
      \bar{g}_{ij} \, (\bar{v}^j)_{,k} + \bar{g}_{ij,k} (\bar{v}^j))
  \right)
  \nonumber\\
  &= g_V V_\mu
    \bar{g}_{ij} (\partial^\mu \varphi^i) (\bar{v}^j)
    -g_V (\partial^\mu V_\mu) \varphi^i \varphi^k (
      \bar{g}_{ij} \, (\bar{v}^j)_{,k} + \bar{g}_{ij,k} (\bar{v}^j))
  \nonumber\\
  & \quad
   +\frac{1}{2} g_V V_\mu (\partial^\mu \varphi^i) \varphi^k \left(
      \bar{g}_{ij} (\bar{v}^j)_{,k} +\bar{g}_{ij,k} (\bar{v}^j)
     -\bar{g}_{kj} (\bar{v}^j)_{,i} -\bar{g}_{kj,i} (\bar{v}^j)
  \right) + \cdots \, .
\end{align}
On the other hand, it is straightforward to show
\begin{eqnarray}
  g_{ij} (v^j)_{;k} - g_{kj} (v^j)_{;i}
  &=& g_{ij} (v^j)_{,k} + g_{ij} \Gamma^j_{kl} v^l
     -g_{kj} (v^j)_{,i} + g_{kj} \Gamma^j_{il} v^l
  \nonumber\\
  &=& g_{ij} (v^j)_{,k} + \dfrac{1}{2} \left[ 
        g_{il,k} + g_{ki,l} - g_{kl,i} \right] v^l
  \nonumber\\
  & & 
     -g_{kj} (v^j)_{,i} - \dfrac{1}{2} \left[ 
        g_{kl,i} + g_{ik,l} - g_{il,k} \right] v^l
  \nonumber\\
  &=& g_{ij} (v^j)_{,k} + g_{il,k} (v^l)
     -g_{kj} (v^j)_{,i} - g_{kl,i} (v^l) \, .
\end{eqnarray}
Here the Affine connection $\Gamma^j_{kl}$ is defined by
\begin{equation}
  \Gamma^j_{kl} 
  = \dfrac{1}{2} g^{jm} \left(
      g_{ml,k} + g_{km,l} - g_{kl,m}
    \right) \, .
\end{equation}
It is now easy to see
\begin{align}
  {\cal L}'_{V\phi} 
  &= g_V V_\mu
    \bar{g}_{ij} (\partial^\mu \varphi^i) (\bar{v}^j)
   -g_V (\partial^\mu V_\mu) \varphi^i \varphi^k (
      \bar{g}_{ij} \, (\bar{v}^j)_{,k} + \bar{g}_{ij,k} (\bar{v}^j))
  \nonumber\\
  & \quad   
   +\frac{1}{2} g_V V_\mu (\partial^\mu \varphi^i) \varphi^k \left(
      \bar{g}_{ij} (\bar{v}^j)_{;k}
     -\bar{g}_{kj} (\bar{v}^j)_{;i}
  \right) + \cdots \, .
\label{eq:action3}
\end{align}
Thanks to the fermion current conservation, the term proportional 
to $\partial^\mu V_\mu$ does not contribute to the 
$f\bar{f} \to V_\mu \to \varphi^i \varphi^j$ amplitude.
It is now easy to show 
\begin{equation}
  {\cal M}(f\bar{f} \to V_\mu \to \varphi^i \varphi^j) \propto 
   \left( \bar{g}_{ik} (\bar{v}^k)_{; j} - \bar{g}_{jk} (\bar{v}^k)_{; i} \right)
   \dfrac{g_V^2 \delta^{ab}}{s-M_V^2} \, .
\end{equation}
The first covariant derivative of the Killing vector,
$\bar{g}_{ik} (\bar{v}^k)_{; j}$, thus plays the role of the 
$V_\mu$-$\varphi^i$-$\varphi^j$ interaction vertex in the
$f\bar{f} \to V_\mu \to \varphi^i \varphi^j$ amplitude.


{}


\begin{thebibliography}{99}



\bibitem{Aad:2012tfa}
  G.~Aad {\it et al.}  [ATLAS Collaboration],
  Phys.\ Lett.\ B {\bf 716} (2012) 1
  [arXiv:1207.7214 [hep-ex]].
  
\bibitem{Chatrchyan:2012ufa}
  S.~Chatrchyan {\it et al.}  [CMS Collaboration],
  Phys.\ Lett.\ B {\bf 716} (2012) 30
  [arXiv:1207.7235 [hep-ex]].
  


\bibitem{Haber:1978jt}
  H.~E.~Haber, G.~L.~Kane and T.~Sterling,
  Nucl.\ Phys.\ B {\bf 161} (1979) 493.
  doi:10.1016/0550-3213(79)90225-6

\bibitem{Deshpande:1977rw}
  N.~G.~Deshpande and E.~Ma,
  Phys.\ Rev.\ D {\bf 18} (1978) 2574.
  doi:10.1103/PhysRevD.18.2574

\bibitem{Georgi:1978xz}
  H.~Georgi,
  Hadronic J.\  {\bf 1} (1978) 155.

\bibitem{Donoghue:1978cj}
  J.~F.~Donoghue and L.~F.~Li,
  Phys.\ Rev.\ D {\bf 19} (1979) 945.
  doi:10.1103/PhysRevD.19.945
  
\bibitem{Abbott:1979dt}
  L.~F.~Abbott, P.~Sikivie and M.~B.~Wise,
  Phys.\ Rev.\ D {\bf 21} (1980) 1393.
  doi:10.1103/PhysRevD.21.1393

\bibitem{McWilliams:1980kj}
  B.~McWilliams and L.~F.~Li,
  Nucl.\ Phys.\ B {\bf 179} (1981) 62.
  doi:10.1016/0550-3213(81)90249-2

\bibitem{Gunion:1984yn}
  J.~F.~Gunion and H.~E.~Haber,
  Nucl.\ Phys.\ B {\bf 272} (1986) 1
   Erratum: [Nucl.\ Phys.\ B {\bf 402} (1993) 567].
  doi:10.1016/0550-3213(86)90340-8, 10.1016/0550-3213(93)90653-7



\bibitem{Branco:2011iw}
  G.~C.~Branco, P.~M.~Ferreira, L.~Lavoura, M.~N.~Rebelo, M.~Sher and J.~P.~Silva,
  Phys.\ Rept.\  {\bf 516} (2012) 1
  [arXiv:1106.0034 [hep-ph]].


\bibitem{Cheon:2012rh}
  H.~S.~Cheon and S.~K.~Kang,
  JHEP {\bf 1309} (2013) 085
  [arXiv:1207.1083 [hep-ph]].
  
\bibitem{Craig:2012vn}
  N.~Craig and S.~Thomas,
  JHEP {\bf 1211} (2012) 083
  [arXiv:1207.4835 [hep-ph]].
  
  
\bibitem{Chang:2012ve}
  S.~Chang, S.~K.~Kang, J.~P.~Lee, K.~Y.~Lee, S.~C.~Park and J.~Song,
  JHEP {\bf 1305} (2013) 075
  [arXiv:1210.3439 [hep-ph]].
  
  
\bibitem{Bai:2012ex}
  Y.~Bai, V.~Barger, L.~L.~Everett and G.~Shaughnessy,
  Phys.\ Rev.\ D {\bf 87} (2013) 11,  115013
  [arXiv:1210.4922 [hep-ph]].
  
\bibitem{Ferreira:2012nv}
  P.~M.~Ferreira, R.~Santos, H.~E.~Haber and J.~P.~Silva,
  Phys.\ Rev.\ D {\bf 87} (2013) 5,  055009
  [arXiv:1211.3131 [hep-ph]].
  
\bibitem{Chang:2012zf}
  J.~Chang, K.~Cheung, P.~Y.~Tseng and T.~C.~Yuan,
  Phys.\ Rev.\ D {\bf 87} (2013) 3,  035008
  [arXiv:1211.3849 [hep-ph]].
  
  
\bibitem{Chen:2013kt}
  C.~Y.~Chen and S.~Dawson,
  Phys.\ Rev.\ D {\bf 87} (2013) 5,  055016
  [arXiv:1301.0309 [hep-ph]].
  
              
\bibitem{Celis:2013rcs}
  A.~Celis, V.~Ilisie and A.~Pich,
  JHEP {\bf 1307} (2013) 053
  [arXiv:1302.4022 [hep-ph]].

\bibitem{Grinstein:2013npa}
  B.~Grinstein and P.~Uttayarat,
  JHEP {\bf 1306} (2013) 094
   [Erratum-ibid.\  {\bf 1309} (2013) 110]
  [arXiv:1304.0028 [hep-ph]].
  
  

\bibitem{Chen:2013rba}
  C.~Y.~Chen, S.~Dawson and M.~Sher,
  Phys.\ Rev.\ D {\bf 88} (2013) 015018
  [arXiv:1305.1624 [hep-ph]].
      
\bibitem{Craig:2013hca}
  N.~Craig, J.~Galloway and S.~Thomas,
  arXiv:1305.2424 [hep-ph].
  
    
\bibitem{Kanemura:2013eja}
  S.~Kanemura, K.~Tsumura and H.~Yokoya,
  Phys.\ Rev.\ D {\bf 88} (2013) 5,  055010
  [arXiv:1305.5424 [hep-ph]].
  
  
\bibitem{Ferreira:2014naa}
  P.~M.~Ferreira, J.~F.~Gunion, H.~E.~Haber and R.~Santos,
  Phys.\ Rev.\ D {\bf 89} (2014) 115003
  [arXiv:1403.4736 [hep-ph]].
      
\bibitem{Kanemura:2014bqa}
  S.~Kanemura, K.~Tsumura, K.~Yagyu and H.~Yokoya,
  Phys.\ Rev.\ D {\bf 90} (2014) 7,  075001
  [arXiv:1406.3294 [hep-ph]].


\bibitem{Kaplan:1983fs}
  D.~B.~Kaplan and H.~Georgi,
  Phys.\ Lett.\  {\bf 136B} (1984) 183.
  doi:10.1016/0370-2693(84)91177-8

\bibitem{Kaplan:1983sm}
  D.~B.~Kaplan, H.~Georgi and S.~Dimopoulos,
  Phys.\ Lett.\  {\bf 136B} (1984) 187.
  doi:10.1016/0370-2693(84)91178-X

\bibitem{Georgi:1984ef}
  H.~Georgi, D.~B.~Kaplan and P.~Galison,
  Phys.\ Lett.\  {\bf 143B} (1984) 152.
  doi:10.1016/0370-2693(84)90823-2

\bibitem{Georgi:1984af}
  H.~Georgi and D.~B.~Kaplan,
  Phys.\ Lett.\  {\bf 145B} (1984) 216.
  doi:10.1016/0370-2693(84)90341-1

\bibitem{Dugan:1984hq}
  M.~J.~Dugan, H.~Georgi and D.~B.~Kaplan,
  Nucl.\ Phys.\ B {\bf 254} (1985) 299.
  doi:10.1016/0550-3213(85)90221-4
  
\bibitem{Contino:2003ve}
  R.~Contino, Y.~Nomura and A.~Pomarol,
  Nucl.\ Phys.\ B {\bf 671} (2003) 148
  doi:10.1016/j.nuclphysb.2003.08.027
  [hep-ph/0306259].


\bibitem{Agashe:2004rs}
  K.~Agashe, R.~Contino and A.~Pomarol,
  Nucl.\ Phys.\ B {\bf 719} (2005) 165
  doi:10.1016/j.nuclphysb.2005.04.035
  [hep-ph/0412089].


\bibitem{Mrazek:2011iu}
  J.~Mrazek, A.~Pomarol, R.~Rattazzi, M.~Redi, J.~Serra and A.~Wulzer,
  Nucl.\ Phys.\ B {\bf 853} (2011) 1
  doi:10.1016/j.nuclphysb.2011.07.008
  [arXiv:1105.5403 [hep-ph]].
  
\bibitem{DeCurtis:2018iqd}
  S.~De Curtis, L.~Delle Rose, S.~Moretti and K.~Yagyu,
  Phys.\ Lett.\ B {\bf 786} (2018) 189
  doi:10.1016/j.physletb.2018.09.042
  [arXiv:1803.01865 [hep-ph]].

\bibitem{DeCurtis:2018zvh}
  S.~De Curtis, L.~Delle Rose, S.~Moretti and K.~Yagyu,
  JHEP {\bf 1812} (2018) 051
  doi:10.1007/JHEP12(2018)051
  [arXiv:1810.06465 [hep-ph]].



\bibitem{Georgi:1985nv}
  H.~Georgi and M.~Machacek,
  Nucl.\ Phys.\ B {\bf 262} (1985) 463.

\bibitem{Chanowitz:1985ug}
  M.~S.~Chanowitz and M.~Golden,
  Phys.\ Lett.\ B {\bf 165} (1985) 105.

\bibitem{Gunion:1989ci}
  J.~F.~Gunion, R.~Vega and J.~Wudka,
  Phys.\ Rev.\ D {\bf 42} (1990) 1673.

\bibitem{Gunion:1990dt}
  J.~F.~Gunion, R.~Vega and J.~Wudka,
  Phys.\ Rev.\ D {\bf 43} (1991) 2322.




\bibitem{Buchmuller:1985jz}
  W.~Buchmuller and D.~Wyler,
  Nucl.\ Phys.\ B {\bf 268} (1986) 621.
    
  
\bibitem{De Rujula:1991se}
  A.~De Rujula, M.~B.~Gavela, P.~Hernandez and E.~Masso,
  Nucl.\ Phys.\ B {\bf 384} (1992) 3.

\bibitem{Hagiwara:1992eh}
  K.~Hagiwara, S.~Ishihara, R.~Szalapski and D.~Zeppenfeld,
  Phys.\ Lett.\ B {\bf 283} (1992) 353.

\bibitem{Hagiwara:1993ck}
  K.~Hagiwara, S.~Ishihara, R.~Szalapski and D.~Zeppenfeld,
  Phys.\ Rev.\ D {\bf 48} (1993) 2182.

\bibitem{Hagiwara:1993qt}
  K.~Hagiwara, R.~Szalapski and D.~Zeppenfeld,
  Phys.\ Lett.\ B {\bf 318} (1993) 155
  [hep-ph/9308347].

\bibitem{Alam:1997nk}
  S.~Alam, S.~Dawson and R.~Szalapski,
  Phys.\ Rev.\ D {\bf 57} (1998) 1577
  [hep-ph/9706542].

    
\bibitem{Barger:2003rs}
  V.~Barger, T.~Han, P.~Langacker, B.~McElrath and P.~Zerwas,
  Phys.\ Rev.\ D {\bf 67} (2003) 115001
  [hep-ph/0301097].
    
\bibitem{Kanemura:2008ub}
  S.~Kanemura and K.~Tsumura,
  Eur.\ Phys.\ J.\ C {\bf 63} (2009) 11
  [arXiv:0810.0433 [hep-ph]].
 
 
\bibitem{Grzadkowski:2010es}
  B.~Grzadkowski, M.~Iskrzynski, M.~Misiak and J.~Rosiek,
  JHEP {\bf 1010} (2010) 085
  [arXiv:1008.4884 [hep-ph]].
  
\bibitem{Corbett:2012dm}
  T.~Corbett, O.~J.~P.~Eboli, J.~Gonzalez-Fraile and M.~C.~Gonzalez-Garcia,
  Phys.\ Rev.\ D {\bf 86} (2012) 075013
  [arXiv:1207.1344 [hep-ph]].


\bibitem{Corbett:2012ja}
  T.~Corbett, O.~J.~P.~Eboli, J.~Gonzalez-Fraile and M.~C.~Gonzalez-Garcia,
  Phys.\ Rev.\ D {\bf 87} (2013) 015022
  [arXiv:1211.4580 [hep-ph]].
 
\bibitem{Grojean:2013kd}
  C.~Grojean, E.~E.~Jenkins, A.~V.~Manohar and M.~Trott,
  JHEP {\bf 1304} (2013) 016
  [arXiv:1301.2588 [hep-ph]].
 
\bibitem{Elias-Miro:2013gya}
  J.~Elias-Mir\'{o}, J.~R.~Espinosa, E.~Masso and A.~Pomarol,
  JHEP {\bf 1308} (2013) 033
  [arXiv:1302.5661 [hep-ph]].
 
\bibitem{Corbett:2013pja}
  T.~Corbett, O.~J.~P.~Eboli, J.~Gonzalez-Fraile and M.~C.~Gonzalez-Garcia,
  Phys.\ Rev.\ Lett.\  {\bf 111} (2013) 1,  011801
  [arXiv:1304.1151 [hep-ph]].

\bibitem{Mebane:2013cra}
  H.~Mebane, N.~Greiner, C.~Zhang and S.~Willenbrock,
  Phys.\ Lett.\ B {\bf 724} (2013) 259
  [arXiv:1304.1789 [hep-ph]].
     
\bibitem{Belanger:2013xza}
  G.~Belanger, B.~Dumont, U.~Ellwanger, J.~F.~Gunion and S.~Kraml,
  Phys.\ Rev.\ D {\bf 88} (2013) 075008
  [arXiv:1306.2941 [hep-ph]].
 
\bibitem{Elias-Miro:2013mua}
  J.~Elias-Mir\'{o}, J.~R.~Espinosa, E.~Masso and A.~Pomarol,
  JHEP {\bf 1311} (2013) 066
  [arXiv:1308.1879 [hep-ph]].
  
\bibitem{Lopez-Val:2013yba}
  D.~L\'{o}pez-Val, T.~Plehn and M.~Rauch,
  JHEP {\bf 1310} (2013) 134
  [arXiv:1308.1979 [hep-ph]].
 
\bibitem{Jenkins:2013zja}
  E.~E.~Jenkins, A.~V.~Manohar and M.~Trott,
  JHEP {\bf 1310} (2013) 087
  [arXiv:1308.2627 [hep-ph]].
  
\bibitem{Boos:2013mqa}
  E.~Boos, V.~Bunichev, M.~Dubinin and Y.~Kurihara,
  Phys.\ Rev.\ D {\bf 89} (2014) 3,  035001
  [arXiv:1309.5410 [hep-ph]].
   

\bibitem{Jenkins:2013wua}
  E.~E.~Jenkins, A.~V.~Manohar and M.~Trott,
  JHEP {\bf 1401} (2014) 035
  [arXiv:1310.4838 [hep-ph]].

\bibitem{Alonso:2013hga}
  R.~Alonso, E.~E.~Jenkins, A.~V.~Manohar and M.~Trott,
  JHEP {\bf 1404} (2014) 159
  [arXiv:1312.2014 [hep-ph]].
    

\bibitem{Ellis:2014dva}
  J.~Ellis, V.~Sanz and T.~You,
  JHEP {\bf 1407} (2014) 036
  doi:10.1007/JHEP07(2014)036
  [arXiv:1404.3667 [hep-ph]].

\bibitem{Ellis:2014jta}
  J.~Ellis, V.~Sanz and T.~You,
  JHEP {\bf 1503} (2015) 157
  doi:10.1007/JHEP03(2015)157
  [arXiv:1410.7703 [hep-ph]].

\bibitem{Falkowski:2014tna}
  A.~Falkowski and F.~Riva,
  JHEP {\bf 1502} (2015) 039
  doi:10.1007/JHEP02(2015)039
  [arXiv:1411.0669 [hep-ph]].

\bibitem{Henning:2014wua}
  B.~Henning, X.~Lu and H.~Murayama,
  JHEP {\bf 1601} (2016) 023
  doi:10.1007/JHEP01(2016)023
  [arXiv:1412.1837 [hep-ph]].

\bibitem{Contino:2016jqw}
  R.~Contino, A.~Falkowski, F.~Goertz, C.~Grojean and F.~Riva,
  JHEP {\bf 1607} (2016) 144
  doi:10.1007/JHEP07(2016)144
  [arXiv:1604.06444 [hep-ph]].

\bibitem{Ellis:2018gqa}
  J.~Ellis, C.~W.~Murphy, V.~Sanz and T.~You,
  JHEP {\bf 1806} (2018) 146
  doi:10.1007/JHEP06(2018)146
  [arXiv:1803.03252 [hep-ph]].




\bibitem{Feruglio:1992wf}
  F.~Feruglio,
  Int.\ J.\ Mod.\ Phys.\ A {\bf 8} (1993) 4937
  doi:10.1142/S0217751X93001946
  [hep-ph/9301281].

\bibitem{Burgess:1999ha}
  C.~P.~Burgess, J.~Matias and M.~Pospelov,
  Int.\ J.\ Mod.\ Phys.\ A {\bf 17} (2002) 1841
  doi:10.1142/S0217751X02009813
  [hep-ph/9912459].

\bibitem{Giudice:2007fh}
  G.~F.~Giudice, C.~Grojean, A.~Pomarol and R.~Rattazzi,
  JHEP {\bf 0706} (2007) 045
  [hep-ph/0703164].

\bibitem{Grinstein:2007iv}
  B.~Grinstein and M.~Trott,
  Phys.\ Rev.\ D {\bf 76} (2007) 073002
  [arXiv:0704.1505 [hep-ph]].
 
\bibitem{Azatov:2012bz}
  A.~Azatov, R.~Contino and J.~Galloway,
  JHEP {\bf 1204} (2012) 127
   [Erratum-ibid.\  {\bf 1304} (2013) 140]
  [arXiv:1202.3415 [hep-ph]].
  
\bibitem{Buchalla:2012qq}
  G.~Buchalla and O.~Cata,
  JHEP {\bf 1207} (2012) 101
  [arXiv:1203.6510 [hep-ph]].
  
\bibitem{Alonso:2012px}
  R.~Alonso, M.~B.~Gavela, L.~Merlo, S.~Rigolin and J.~Yepes,
  Phys.\ Lett.\ B {\bf 722} (2013) 330
  [arXiv:1212.3305 [hep-ph]].

\bibitem{Contino:2013kra}
  R.~Contino, M.~Ghezzi, C.~Grojean, M.~Muhlleitner and M.~Spira,
  JHEP {\bf 1307} (2013) 035
  [arXiv:1303.3876 [hep-ph]].

\bibitem{Jenkins:2013fya}
  E.~E.~Jenkins, A.~V.~Manohar and M.~Trott,
  JHEP {\bf 1309} (2013) 063
  [arXiv:1305.0017 [hep-ph]].
      
\bibitem{Buchalla:2013rka}
  G.~Buchalla, O.~Cata and C.~Krause,
  Nucl.\ Phys.\ B {\bf 880} (2014) 552
  [arXiv:1307.5017 [hep-ph]].

\bibitem{Buchalla:2013eza}
  G.~Buchalla, O.~Cat\'{a} and C.~Krause,
  Phys.\ Lett.\ B {\bf 731} (2014) 80
  doi:10.1016/j.physletb.2014.02.015
  [arXiv:1312.5624 [hep-ph]].

\bibitem{Alonso:2014rga}
  R.~Alonso, E.~E.~Jenkins and A.~V.~Manohar,
  arXiv:1409.0868 [hep-ph].

\bibitem{Guo:2015isa}
  F.~K.~Guo, P.~Ruiz-Femen\'{i}a and J.~J.~Sanz-Cillero,
  Phys.\ Rev.\ D {\bf 92} (2015) 074005
  doi:10.1103/PhysRevD.92.074005
  [arXiv:1506.04204 [hep-ph]].

\bibitem{Buchalla:2015qju}
  G.~Buchalla, O.~Cata, A.~Celis and C.~Krause,
  Eur.\ Phys.\ J.\ C {\bf 76} (2016) no.5,  233
  doi:10.1140/epjc/s10052-016-4086-9
  [arXiv:1511.00988 [hep-ph]].

\bibitem{Buchalla:2017jlu}
  G.~Buchalla, O.~Cata, A.~Celis, M.~Knecht and C.~Krause,
  Nucl.\ Phys.\ B {\bf 928} (2018) 93
  doi:10.1016/j.nuclphysb.2018.01.009
  [arXiv:1710.06412 [hep-ph]].

\bibitem{Alonso:2017tdy}
  R.~Alonso, K.~Kanshin and S.~Saa,
  Phys.\ Rev.\ D {\bf 97} (2018) no.3,  035010
  doi:10.1103/PhysRevD.97.035010
  [arXiv:1710.06848 [hep-ph]].

\bibitem{Buchalla:2018yce}
  G.~Buchalla, M.~Capozi, A.~Celis, G.~Heinrich and L.~Scyboz,
  JHEP {\bf 1809} (2018) 057
  doi:10.1007/JHEP09(2018)057
  [arXiv:1806.05162 [hep-ph]].



\bibitem{Appelquist:1980vg}
  T.~Appelquist and C.~W.~Bernard,
  Phys.\ Rev.\ D {\bf 22} (1980) 200.
  doi:10.1103/PhysRevD.22.200
 
\bibitem{Longhitano:1980iz}
  A.~C.~Longhitano,
  Phys.\ Rev.\ D {\bf 22} (1980) 1166.
  doi:10.1103/PhysRevD.22.1166
 
\bibitem{Longhitano:1980tm}
  A.~C.~Longhitano,
  Nucl.\ Phys.\ B {\bf 188} (1981) 118.
  doi:10.1016/0550-3213(81)90109-7
  
\bibitem{Appelquist:1980ae}
  T.~Appelquist and C.~W.~Bernard,
  Phys.\ Rev.\ D {\bf 23} (1981) 425.
  doi:10.1103/PhysRevD.23.425
  
\bibitem{Appelquist:1993ka}
  T.~Appelquist and G.~H.~Wu,
  Phys.\ Rev.\ D {\bf 48} (1993) 3235
  doi:10.1103/PhysRevD.48.3235
  [hep-ph/9304240].

\bibitem{Appelquist:1994qz}
  T.~Appelquist and G.~H.~Wu,
  Phys.\ Rev.\ D {\bf 51} (1995) 240
  doi:10.1103/PhysRevD.51.240
  [hep-ph/9406416].

\bibitem{Coleman:1969sm}
  S.~R.~Coleman, J.~Wess and B.~Zumino,
  Phys.\ Rev.\  {\bf 177} (1969) 2239.
  doi:10.1103/PhysRev.177.2239
 
\bibitem{Callan:1969sn}
  C.~G.~Callan, Jr., S.~R.~Coleman, J.~Wess and B.~Zumino,
  Phys.\ Rev.\  {\bf 177} (1969) 2247.
  doi:10.1103/PhysRev.177.2247
 
\bibitem{Bando:1987br}
  M.~Bando, T.~Kugo and K.~Yamawaki,
  Phys.\ Rept.\  {\bf 164} (1988) 217.
  doi:10.1016/0370-1573(88)90019-1
 


\bibitem{Gunion:1990kf}
  J.~F.~Gunion, H.~E.~Haber and J.~Wudka,
  Phys.\ Rev.\ D {\bf 43} (1991) 904.

\bibitem{Csaki:2003dt}
  C.~Csaki, C.~Grojean, H.~Murayama, L.~Pilo and J.~Terning,
  Phys.\ Rev.\ D {\bf 69} (2004) 055006
  [hep-ph/0305237].

\bibitem{SekharChivukula:2008mj}
  R.~S.~Chivukula, H.~J.~He, M.~Kurachi, E.~H.~Simmons and M.~Tanabashi,
  Phys.\ Rev.\ D {\bf 78} (2008) 095003
  [arXiv:0808.1682 [hep-ph]].



\bibitem{Alonso:2015fsp}
  R.~Alonso, E.~E.~Jenkins and A.~V.~Manohar,
  Phys.\ Lett.\ B {\bf 754} (2016) 335
  doi:10.1016/j.physletb.2016.01.041
  [arXiv:1511.00724 [hep-ph]].

\bibitem{Alonso:2016oah}
  R.~Alonso, E.~E.~Jenkins and A.~V.~Manohar,
  JHEP {\bf 1608} (2016) 101
  doi:10.1007/JHEP08(2016)101
  [arXiv:1605.03602 [hep-ph]].


\bibitem{Peskin:1990zt}
  M.~E.~Peskin and T.~Takeuchi,
  Phys.\ Rev.\ Lett.\  {\bf 65} (1990) 964.



\bibitem{Nagai:2014cua}
  R.~Nagai, M.~Tanabashi and K.~Tsumura,
  Phys.\ Rev.\ D {\bf 91} (2015) no.3,  034030
  doi:10.1103/PhysRevD.91.034030
  [arXiv:1409.1709 [hep-ph]].
  
  
\bibitem{Fujimori:2015wda}
  T.~Fujimori, T.~Inami, K.~Izumi and T.~Kitamura,
  Phys.\ Rev.\ D {\bf 91} (2015) no.12,  125007
  doi:10.1103/PhysRevD.91.125007
  [arXiv:1502.01820 [hep-th]].
  
\bibitem{Fujimori:2015mea}
  T.~Fujimori, T.~Inami, K.~Izumi and T.~Kitamura,
  PTEP {\bf 2016} (2016) no.1,  013B08
  doi:10.1093/ptep/ptv185
  [arXiv:1510.07237 [hep-th]].
  
\bibitem{Fujimori:2016rrc}
  T.~Fujimori, T.~Inami, K.~Izumi and T.~Kitamura,
  doi:10.1142/9789813203952 \_0054
  arXiv:1601.06470 [hep-th].
  
\bibitem{Abe:2017abx}
  Y.~Abe, T.~Inami, K.~Izumi and T.~Kitamura,
  PTEP {\bf 2018} (2018) no.3,  031E01
  doi:10.1093/ptep/pty010
  [arXiv:1712.06305 [hep-th]].
  
\bibitem{Abe:2018rwb}
  Y.~Abe, T.~Inami, K.~Izumi, T.~Kitamura and T.~Noumi,
  arXiv:1805.00262 [hep-th].

    

  
\bibitem{Cacciapaglia:2004rb}
  G.~Cacciapaglia, C.~Csaki, C.~Grojean and J.~Terning,
  Phys.\ Rev.\ D {\bf 71} (2005) 035015
  doi:10.1103/PhysRevD.71.035015
  [hep-ph/0409126].


\bibitem{Foadi:2004ps}
  R.~Foadi, S.~Gopalakrishna and C.~Schmidt,
  Phys.\ Lett.\ B {\bf 606} (2005) 157
  doi:10.1016/j.physletb.2004.11.055
  [hep-ph/0409266].

\bibitem{Chivukula:2005bn}
  R.~S.~Chivukula, E.~H.~Simmons, H.~J.~He, M.~Kurachi and M.~Tanabashi,
  Phys.\ Rev.\ D {\bf 71} (2005) 115001
  doi:10.1103/PhysRevD.71.115001
  [hep-ph/0502162].

\bibitem{Casalbuoni:2005rs}
  R.~Casalbuoni, S.~De Curtis, D.~Dolce and D.~Dominici,
  Phys.\ Rev.\ D {\bf 71} (2005) 075015
  doi:10.1103/PhysRevD.71.075015
  [hep-ph/0502209].

\bibitem{Chivukula:2005xm}
  R.~S.~Chivukula, E.~H.~Simmons, H.~J.~He, M.~Kurachi and M.~Tanabashi,
  Phys.\ Rev.\ D {\bf 72} (2005) 015008
  doi:10.1103/PhysRevD.72.015008
  [hep-ph/0504114].

\bibitem{Cacciapaglia:2005pa}
  G.~Cacciapaglia, C.~Csaki, C.~Grojean, M.~Reece and J.~Terning,
  Phys.\ Rev.\ D {\bf 72} (2005) 095018
  doi:10.1103/PhysRevD.72.095018
  [hep-ph/0505001].

\bibitem{Foadi:2005hz}
  R.~Foadi and C.~Schmidt,
  Phys.\ Rev.\ D {\bf 73} (2006) 075011
  doi:10.1103/PhysRevD.73.075011
  [hep-ph/0509071].

\bibitem{Chivukula:2006cg}
  R.~S.~Chivukula, B.~Coleppa, S.~Di Chiara, E.~H.~Simmons, H.~J.~He, M.~Kurachi and M.~Tanabashi,
  Phys.\ Rev.\ D {\bf 74} (2006) 075011
  doi:10.1103/PhysRevD.74.075011
  [hep-ph/0607124].

\bibitem{Abe:2008hb}
  T.~Abe, S.~Matsuzaki and M.~Tanabashi,
  Phys.\ Rev.\ D {\bf 78} (2008) 055020
  doi:10.1103/PhysRevD.78.055020
  [arXiv:0807.2298 [hep-ph]].

\bibitem{Abe:2011sv}
  T.~Abe, R.~S.~Chivukula, E.~H.~Simmons and M.~Tanabashi,
  Phys.\ Rev.\ D {\bf 85} (2012) 035015
  doi:10.1103/PhysRevD.85.035015
  [arXiv:1109.5856 [hep-ph]].


\bibitem{Ecker:1988te}
  G.~Ecker, J.~Gasser, A.~Pich and E.~de Rafael,
  Nucl.\ Phys.\ B {\bf 321} (1989) 311.
  doi:10.1016/0550-3213(89)90346-5


\bibitem{Alboteanu:2008my}
  A.~Alboteanu, W.~Kilian and J.~Reuter,
  JHEP {\bf 0811} (2008) 010
  doi:10.1088/1126-6708/2008/11/010
  [arXiv:0806.4145 [hep-ph]].

\bibitem{deFlorian:2016spz}
  D.~de Florian {\it et al.} [LHC Higgs Cross Section Working Group],
  doi:10.23731/CYRM-2017-002
  arXiv:1610.07922 [hep-ph].


\bibitem{Tanabashi:1993np}
  M.~Tanabashi,
  Phys.\ Lett.\ B {\bf 316} (1993) 534
  doi:10.1016/0370-2693(93)91040-T
  [hep-ph/9306237].

\bibitem{Rosell:2005ai}
  I.~Rosell, P.~Ruiz-Femenia and J.~Portoles,
  JHEP {\bf 0512} (2005) 020
  doi:10.1088/1126-6708/2005/12/020
  [hep-ph/0510041].




 \bibitem{Zhang:2003it}
   B.~Zhang, Y.~-P.~Kuang, H.~-J.~He and C.~P.~Yuan,
   Phys.\ Rev.\ D {\bf 67} (2003) 114024
   [hep-ph/0303048].

\bibitem{Chang:2013aya}
  J.~Chang, K.~Cheung, C.~-T.~Lu and T.~-C.~Yuan,
  Phys.\ Rev.\ D {\bf 87} (2013) 093005
  [arXiv:1303.6335 [hep-ph]].



\bibitem{Cornwall:1974km}
  J.~M.~Cornwall, D.~N.~Levin and G.~Tiktopoulos,
  Phys.\ Rev.\ D {\bf 10} (1974) 1145
   [Erratum-ibid.\ D {\bf 11} (1975) 972].

\bibitem{Chanowitz:1985hj}
  M.~S.~Chanowitz and M.~K.~Gaillard,
  Nucl.\ Phys.\ B {\bf 261} (1985) 379.
  doi:10.1016/0550-3213(85)90580-2

\bibitem{Gounaris:1986cr}
  G.~J.~Gounaris, R.~Kogerler and H.~Neufeld,
  Phys.\ Rev.\ D {\bf 34} (1986) 3257.
  doi:10.1103/PhysRevD.34.3257

\bibitem{He:1993yd}
  H.~J.~He, Y.~P.~Kuang and X.~y.~Li,
  Phys.\ Rev.\ D {\bf 49} (1994) 4842.
  doi:10.1103/PhysRevD.49.4842

\bibitem{He:1993qa}
  H.~J.~He, Y.~P.~Kuang and X.~y.~Li,
  Phys.\ Lett.\ B {\bf 329} (1994) 278
  doi:10.1016/0370-2693(94)90772-2
  [hep-ph/9403283].


\bibitem{Llewellyn Smith:1973ey}
  C.~H.~Llewellyn Smith,
  Phys.\ Lett.\ B {\bf 46} (1973) 233.

\bibitem{Cornwall:1973tb}
  J.~M.~Cornwall, D.~N.~Levin and G.~Tiktopoulos,
  ``Uniqueness of spontaneously broken gauge theories,''
  Phys.\ Rev.\ Lett.\  {\bf 30} (1973) 1268
   [Erratum-ibid.\  {\bf 31} (1973) 572].

  

\bibitem{Lee:1977eg}
  B.~W.~Lee, C.~Quigg and H.~B.~Thacker,
  Phys.\ Rev.\ D {\bf 16} (1977) 1519.

 

\bibitem{Abe:2015jra}
  T.~Abe, R.~Nagai, S.~Okawa and M.~Tanabashi,
  Phys.\ Rev.\ D {\bf 92} (2015) no.5,  055016
  doi:10.1103/PhysRevD.92.055016
  [arXiv:1507.01185 [hep-ph]].

\bibitem{Abe:2016fjs}
  T.~Abe and R.~Nagai,
  Phys.\ Rev.\ D {\bf 95} (2017) no.7,  075022
  doi:10.1103/PhysRevD.95.075022
  [arXiv:1607.03706 [hep-ph]].

\bibitem{Veltman:1977kh}
  M.~J.~G.~Veltman,
  Nucl.\ Phys.\ B {\bf 123} (1977) 89.
  doi:10.1016/0550-3213(77)90342-X



\bibitem{Honerkamp:1971sh}
  J.~Honerkamp,
  Nucl.\ Phys.\ B {\bf 36} (1972) 130.
  doi:10.1016/0550-3213(72)90299-4


\bibitem{Abbott:1980hw}
  L.~F.~Abbott,
  Nucl.\ Phys.\ B {\bf 185} (1981) 189.
  doi:10.1016/0550-3213(81)90371-0
 
\bibitem{AlvarezGaume:1981hn}
  L.~Alvarez-Gaume, D.~Z.~Freedman and S.~Mukhi,
  Annals Phys.\  {\bf 134} (1981) 85.
  doi:10.1016/0003-4916(81)90006-3
 
\bibitem{Boulware:1981ns}
  D.~G.~Boulware and L.~S.~Brown,
  Annals Phys.\  {\bf 138} (1982) 392.
  doi:10.1016/0003-4916(82)90192-0
  
\bibitem{Howe:1986vm}
  P.~S.~Howe, G.~Papadopoulos and K.~S.~Stelle,
  Nucl.\ Phys.\ B {\bf 296} (1988) 26.
  doi:10.1016/0550-3213(88)90379-3

\bibitem{Fabbrichesi:2010xy}
  M.~Fabbrichesi, R.~Percacci, A.~Tonero and O.~Zanusso,
  Phys.\ Rev.\ D {\bf 83} (2011) 025016
  doi:10.1103/PhysRevD.83.025016
  [arXiv:1010.0912 [hep-ph]].


\bibitem{Muller:1997zk}
  U.~Muller, C.~Schubert and A.~M.~E.~van de Ven,
  Gen.\ Rel.\ Grav.\  {\bf 31} (1999) 1759
  doi:10.1023/A:1026718301634
  [gr-qc/9712092].
  
\bibitem{Hatzinikitas:2000xe}
  A.~Hatzinikitas,
  hep-th/0001078.
  
\bibitem{Cheung:2017pzi}
  C.~Cheung,
  doi:10.1142/9789813233348\_0008
  arXiv:1708.03872 [hep-ph].

\bibitem{Asribekov:1962tgp}
  V.~E.~Asribekov,
  J.\ Exp.\ Theor.\ Phys.\  {\bf 15} (1962) no.2,  394.


 
\end{thebibliography}
\end{document}